\documentclass[11pt]{article}
\usepackage[round,authoryear,semicolon]{natbib}
\usepackage{graphicx} 
\usepackage[rightcaption]{sidecap}
\usepackage[margin=1in]{geometry}
\usepackage{graphicx}
\usepackage{authblk}
\usepackage{color}
\usepackage{soul}	
\usepackage{hyperref}
\usepackage{amssymb}
\usepackage{amsmath}
\usepackage{subfigure}
\usepackage[footnotesize,bf]{caption}
\usepackage{url}
\usepackage{ulem}
\hypersetup{pdfpagelayout=OneColumn,colorlinks=true,urlcolor=blue,citecolor=blue,linkcolor=blue}
\usepackage{footnote}
\usepackage{bm}
\usepackage{parskip}
\usepackage{cleveref}
\usepackage{adjustbox}
\usepackage{algpseudocode}
\usepackage[]{algorithm}
\usepackage{titlesec}
\usepackage{multirow}
\title{Site characterization at Treasure Island and Delaney Park downhole arrays by heterogeneous data assimilation}
\author[1]{Elnaz Seylabi\thanks{Corresponding author (\url{elnaze@unr.edu})}}
\author[2]{Mohamad M. Hallal}
\author[3]{Brady R. Cox}
\affil[1]{Civil and Environmental Engineering Department, University of Nevada, Reno, USA}
\affil[2]{Civil, Architectural, and Environmental Engineering Department, University of Texas at Austin, USA}
\affil[3]{Civil and Environmental Engineering Department, Utah State University, Utah, USA}
\date{}

\begin{document}
\maketitle
\begin{abstract}
This article extends a recently proposed heterogeneous data assimilation technique for site characterization to estimate compression and shear wave velocity ($V_p$ and $V_s$, respectively) and damping at Treasure Island and Delaney Park downhole arrays. The adopted method is based on the joint inversion of earthquake acceleration time series and experimental surface wave dispersion data, and including physical constraints to improve the inverse problem's well-posedness. We first use synthetic data at these two sites to refine the proposed approach and then apply the refined algorithm to real data sets available at the Treasure Island and Delaney Park downhole arrays. The joint inversion results show that the estimated $V_s$ and $V_p$ profiles are in very good agreement with measured profiles at these two sites. Our synthetic and real data experiment results suggest that $V_p$ estimation from inversion at downhole arrays can be improved by integrating the water table depth information or the higher modes of the Rayleigh wave dispersion data. Depending on the site complexity, water table information can also help reduce uncertainties associated with damping estimation. In the last part of this article, we compare the performance of the inverted profiles to other methods used to incorporate spatial variability and wave scattering effects in 1D ground response analysis (GRA). The comparisons show that the joint inversion-based $V_s$ and $V_p$ profiles and damping ratios estimated in this article can effectively integrate the effects of spatial variability and wave scattering into 1D GRAs, especially at the Delaney Park downhole array, which is classified as a poorly modeled site using traditional 1D GRA.
\end{abstract}

\section{Introduction}
Downhole arrays have always been considered as testbeds for validating one-dimensional (1D) ground response analysis (GRA) methods, as they minimize the uncertainty associated with source and path effects influencing site response. However, depending on the site complexity, such as spatial variability of subsurface layers and three-dimensional wave scattering effects, 1D GRAs using 1D site characterization results, which are essentially point measurements and incapable of capturing the site's complexity, generally fail to reproduce the recorded ground motions and the site's empirical transfer function (ETF). For example, based on a synthesis of results from five major research studies which investigated a total of more than 600 downhole array sites, \cite{hallal2021comparison} found that on average, approximately 50\% are poorly modeled using 1D GRAs using a single shear wave velocity ($V_s$) profile derived from invasive site characterization methods. Therefore, the goal of this study is to explore whether 1D GRAs can be improved by using noninvasive site characterization methods over a large area to better capture the representative average subsurface properties influencing site response.

The effects of spatial variability and wave scattering on site response have been widely investigated in the past and most researchers agree that oversimplifications in 1D GRAs, particularly the assumptions of isotropic, homogeneous layers of infinite lateral extent, and ignoring geometric energy loss due to three dimensional wave scattering, greatly limit the efficacy of this standard approach in practice \citep{afshari2019insights, Hallal2021HVPart2, hallal2021comparison, tao2019insights, pilz2019does, thompson2012taxonomy}. Specifically, these researchers have noted that 1D GRAs based on a single $V_s$ profile and laboratory-based estimates of small-strain damping significantly overestimate the site response when using small-strain ground motions. This overestimation of site response has been attributed to wave scattering effects inherent to field-scale problems, which cannot be captured by traditional 1D GRAs using a single $V_s$ profile and laboratory-based estimates of damping in a small specimen.

To address these limitations, several approaches have been proposed in recent years to try and incorporate spatial variability and wave scattering effects through various modifications to 1D GRAs. For example, researchers have recommended modifying damping values to model the apparent energy dissipation and wave scattering \citep{tao2019insights, afshari2019insights}. Alternatively, others have suggested performing numerous standard 1D GRAs using variable small-strain $V_s$ profiles that are meant to capture spatial variability across the site, and then averaging their results. These variable $V_s$ profiles are mostly commonly developed in practice through stochastic $V_s$ randomization models \citep{toro1995}. However, developing a set of site-specific and representative $V_s$ profiles to use in these spatially averaged approaches continues to be a major challenge. \cite{Hallal2021HVPart1} argued that spatial variability is site-specific and difficult to represent by purely stochastic models based on generic parameters derived for other sites. Consequently, more research is required to develop feasible approaches that rely on site-specific estimates of material properties and spatial variability. 

 Recently, \cite{seylabi2020site} utilized a data assimilation technique based on the ensemble Kalman inversion to estimate $V_s$ and damping at the Garner Valley downhole array. Their results suggest that the joint inversion of acceleration time series and dispersion data may produce a composite 1D representation of the site that can effectively incorporate spatial variability and wave scattering effects into site response and consequently, allow for better modeling of the site's ETF. In this article, we extend this implementation to the Treasure Island and Delaney Park downhole arrays to determine whether it is possible to improve the estimation of $V_p$, $V_s$, and damping such that the resulting 1D representation of the subsurface can better reproduce the effects of spatial variability influencing site response. It should be noted that both of these downhole arrays have been extensively studied in the past and most studies concurred on higher site complexity at Delaney Park downhole array and the deficiency of different proposals for modeling spatial variability effects in 1D GRAs \citep{tao2019insights, hallal2021comparison, Hallal2021HVPart2}.
 
 In the remainder of this article, we first summarize the methodology we use for joint inversion of earthquake acceleration time series and surface wave dispersion data to estimate $V_p$, $V_s$, and damping at these two downhole arrays. We then use synthetic data at these two sites to refine the proposed methodology without real data noise and modeling errors. Third, we use real data (i.e., recorded ground motions and measured dispersion data) at Treasure Island and Delaney Park downhole arrays to perform site characterization using the proposed methodology. Lastly, we use the results from our joint inversions in 1D small strain GRAs and discuss the findings of this study in conjunction with previous studies conducted at these two sites.

\section{Methodology}
In this section, we present the inverse problem formulation we use for site characterization at downhole arrays, mainly adopted from the one proposed by \cite{seylabi2020site}. To this end, we model the site as a horizontally layered soil on elastic bedrock and the main objective is to use available observational data sets at the site, that can be different in type and therefore heterogeneous, to estimate P and S wave velocity and small-strain damping at each layer. In this paper, we work with two data sets: the discrete earthquake acceleration time series recorded by the array instruments at different depths ($y_1$), and the experimental surface wave dispersion data that depicts discrete phase velocity values of Rayleigh waves as a function of frequency ($y_2$). If we combine these two data sets, we consider the inverse problem of finding $u$ from data sets $y_1$ and $y_2$ such that:
\begin{equation}
\begin{bmatrix}
y_1\\y_2
\end{bmatrix}
=
\begin{bmatrix}
G_1(u)\\
G_2(u)
\end{bmatrix}
+
\begin{bmatrix}
\eta_1\\\eta_2
\end{bmatrix}
\rightarrow y = G(u)+\eta, \quad 
\Gamma = \begin{bmatrix}
\Gamma_1 & 0 \\ 0 & \Gamma_2
\end{bmatrix}\,.
\end{equation}
We note here that $u = \{u_1,u_2,\dots,u_k\}$ is an array of $k$ unknown/uncertain parameters (P and S wave velocity at each layer and damping) to be estimated, $y_i$ is an array of $m_i$ data points the i$^{th}$ data set has, $\eta_i$ is the noise array represented as independent zero-mean Gaussian noise with covariance matrix $\Gamma_i$ and $G_i$ is a nonlinear function of $u$ (referred to as the forward model) in a form of an array of length $m_i$ that maps the parameter space $u$ to the i$^{th}$ data set $y_i$. We define the covariance matrix of the Gaussian noise as follows:
\begin{equation}
\Gamma_1 = \left[\beta_1\text{diag}(\max|y_1|\bm{1})\right]^2, \quad \Gamma_2 = \left[\beta_2\text{diag}( y_2)\right]^2
\label{eq:gamma}
\end{equation}
where $\beta_1$ and $\beta_2$ determine the noise levels for $y_1$ and $y_2$, and $\bf 1$ is a unit array of size $m_1$. For the two data sets relevant to the problem in hand, the forward models are described below: 
\begin{itemize}
\item[i)] For the theoretical acceleration time series, we consider wave propagation in a horizontally stratified layered soil of total thickness $H$ and shear wave velocity $V_s(z)$ varying with depth $z$. Given an acceleration time series at $z = H$ (i.e., the borehole sensor depth), we compute the soil response numerically using a finite element model and we use the extended Rayleigh damping \citep{phillips2009damping} to capture the nearly frequency independent viscous damping $\xi$ in time domain analyses. We note that $\xi$ is assumed to be constant (i.e., depth-independent), which can be considered as a representative of the average/overall $\xi$.
\item[ii)] For the theoretical dispersion curve we use the transfer matrix approach originally developed by \cite{thomson1950transmission} and \cite{haskell1953dispersion} and later modified by \cite{dunkin1965computation} and \cite{knopoff1964matrix}. This approach requires the solution of an eigenvalue problem, for which we use the well known software Geopsy \citep{wathelet2005array}.
\end{itemize}

To solve the inverse problem involving the two data sets just described, we use a sequential data assimilation method \citep{evensen2009data}  based on the ensemble Kalman inversion \citep{iglesias2013ensemble}, a methodology pioneered in the oil reservoir community \citep{O1,O2}. We provide the details of this algorithm and successful inclusion of a priori knowledge in a form of inequality and equality constraints in \cite{seylabi2020site,albers2019ensemble}. Therefore, for brevity, next we only discuss the main steps of the adopted algorithm formulated in the range of the covariance. In this algorithm, we first generate an initial ensemble of $N$ particles $\{u_0^{(n)}\}_{n=1}^N$ at iteration $j=0$. Considering a horizontally layered soil with $l$ layers, each particle is an array of size $k = 2l+1$ including P and S wave velocity of each layer and a damping ratio. We note that in this study we assume that the damping ratio is depth independent. To generate the initial ensemble of $N$ particles, we consider uniform distribution for each parameter. Then, at each iteration $j$, we use the forward model predictions $\{G(u_j^{(n)})\}_{n=1}^N$ and the observation data $y$ to update these particles sequentially. We have two forward models: $G_1(u_j^{(n)})$ uses the $n$th $V_s$ profiles and damping ratio to perform 1D site response analysis and compute the acceleration time series at different depths. On the other hand, $G_2(u_j^{(n)})$ uses the $n$th $V_p$ and $V_s$ profiles to compute the theoretical dispersion curve. For each particle, we concatenate the forward models' prediction results to form $G(u_j^{(n)})$ and compute the prediction error $y-G(u_j^{n})$. Then, we use the modeling error along with the Kalman gain $C_{j+1}^{uw}(C_{j+1}^{ww}+\Gamma)^{-1}$ to correct each particle as follows:
\begin{equation}
u_{j+1}^{(n)} = u_j^{(n)} + C_{j+1}^{uw}(C_{j+1}^{ww}+\Gamma)^{-1}\left(y-G(u_j^{(n)})\right) \quad \text{for} \quad n = 1, \dots, N \,.
\label{eq:110}
\end{equation}
Matrices $C_{j+1}^{uw}$ and $C_{j+1}^{ww}$ are empirical covariance matrices that can be computed at each iteration based on predictions and the ensemble mean $\bar{u}_{j+1}$ using the following equations.
\begin{equation}
C_{j+1}^{uw} = \frac{1}{N}\sum_{n=1}^{N}(u_j^{(n)}-\bar{u}_{j+1})\otimes(G(u_j^{(n)})-\bar{G}_j)
\label{eq:111}
\end{equation}
\begin{equation}
C_{j+1}^{ww} = \frac{1}{N}\sum_{n=1}^N(G(u_j^{(n)})-\bar{G}_j)\otimes(G(u_j^{(n)})-\bar{G}_j)
\label{eq:112}
\end{equation}
where $\otimes$ stands for the outer product and
\begin{equation}
\bar{u}_{j+1} = \frac{1}{N}\sum_{n=1}^Nu_j^{(n)}, \quad \bar{G}_j = \frac{1}{N}\sum_{n=1}^{N}G(u_j^{(n)}) \,.
\label{eq:mean}
\end{equation}
After updating each particle using the Kalman gain and forward model prediction error, it is possible that the corrected particle does not satisfy the enforced constraint, i.e. $Au_{j+1}^{(n)}> g$. For violating particles, we repeat the previous step and instead of using \eqref{eq:110}, we solve a constrained quadratic programming optimization problem to find a vector $b^{(n)}$ that minimizes the cost function $J_{j,n}(b)$ defined as
\begin{equation}
\label{eq:min}
J_{j,n}(b) := \frac{1}{2}\left|y-G(u_j^{(n)}) - \frac{1}{N}\sum_{m=1}^N b_m \left(G(u_j^{(m)})-\bar{G}_j\right)\right|_\Gamma^2+\frac{1}{2N}\sum_{m=1}^N (b_m)^2
\end{equation}
and subject to
\begin{equation}
\label{eq:constrain}
ABb \leq g-A{u}_{j}^{(n)}, \quad Bb = \frac{1}{N} \sum_{m=1}^N b_m(u_j^{(m)}-\bar{u}_{j+1})\,.
\end{equation}
$|~.~|_\Gamma$ denotes weighted L2-norm where $|v|_{\Gamma}^2 = v^T \Gamma^{-1} v$ and $v^T$ is the transpose of a vector $v$. Then, we use the computed $b^{(n)}$ to update the violating particle $u_j^{(n)}$ as follows:
\begin{equation}
\label{eq:update2}
u_{j+1}^{(n)} = u_j^{(n)}+\frac{1}{N}\sum_{m=1}^{N}b_m^{(n)}(u_j^{(m)}-\bar{u}_{j+1})\,.
\end{equation} 
After updating all particles at iteration $j$, we move to the next iteration. As a result, we iteratively update an ensemble of $V_s$ and $V_p$ profiles and damping ratio values to find optimal profiles that minimize the error between the forward models' prediction and observation data sets including recorded acceleration time series data at different depths and experimental dispersion data.

\section{Downhole Array Sites}
In this article, we implement the methodology described in the previous section to characterize the subsurface properties at two downhole arrays using synthetic and real data. These downhole arrays are the Treasure Island and Delaney Park downhole arrays, for which we provide geologic, ground motion, and dispersion information in the next subsections.

\subsection{Geologic Information}
\begin{figure}[htbp!]
\centering
  \subfigure[]{\includegraphics[width = 0.49\textwidth]{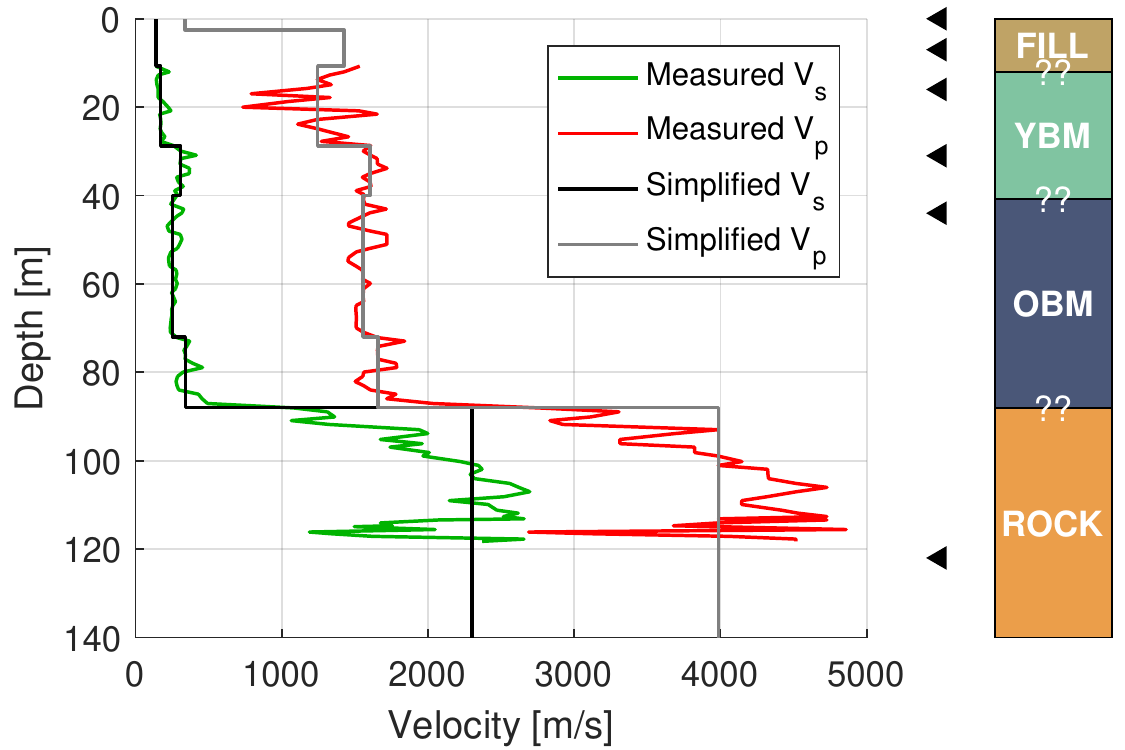}}
  \subfigure[]{\includegraphics[width = 0.49\textwidth]{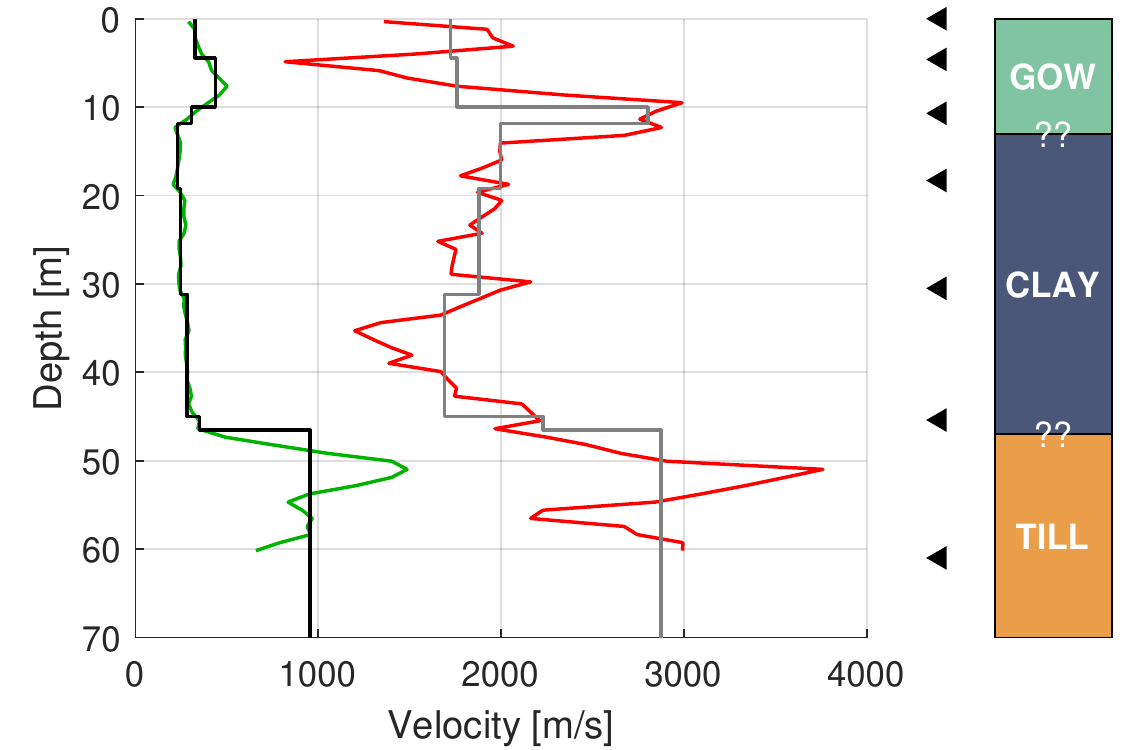}}
   \caption{Measured and simplified invasive $V_s$ and $V_p$ profiles, strong motion sensor locations (black triangle symbols), and simplified 1D stratigraphy at (a) TIDA and (b) DPDA.}
   \label{fig:Velocity_Profiles}
\end{figure}

The Treasure Island Downhole Array (TIDA) site is located in northern California on an artificial hydraulic fill island constructed on a natural sandspit northwest of Yerba Buena Island, an outcropping bedrock island in San  Francisco  Bay  \citep{rollins1994ground}.  Measured  $V_s$  and $V_p$ profiles using  PS  suspension  logging  \citep{graizer2004analysis} along with the simplified stratigraphy are shown in \Cref{fig:Velocity_Profiles}a. Existing data indicates that the subsurface near the downhole array site consists of loose sandy hydraulic fill, Young Bay Mud (YBM), and Old Bay Mud (OBM) overlying bedrock. The bedrock is located at approximately 90 m and is composed of interbedded sandstone, siltstone, and shale of the Franciscan Formation \citep{gibbs1992seismic,pass1994soil}. 
The TIDA site is instrumented with triaxial accelerometers at the surface and throughout the soil profile  into  the  bedrock at depths of 0, 7, 16, 31, 44, and 122 m, as shown using black triangle symbols in \Cref{fig:Velocity_Profiles}a. The water table is reported at a depth shallower than approximately 4 m \citep{foerster2007nonlinear}.

The Delaney Park Downhole Array (DPDA) site  is  located  in  downtown  Anchorage, Alaska in  Delaney  Park. $V_s$ and $V_p$  profiles measured using seismic downhole \citep{thornley2019situ} are shown in \Cref{fig:Velocity_Profiles}b. Geologic data indicates that the subsurface near the downhole array site consists of the glacial outwash (GOW), followed by clay from the Bootlegger Cove Formation overlying glacial till \citep{combellick1999simplified}. The glacial till is located at approximately 47 m and consists of very dense sand and gravel. The strong motion sensors currently available at DPDA site are located at depths of 0, 4.6, 10.7, 18.3, 30.5, 45.4, and 61 m, as shown in \Cref{fig:Velocity_Profiles}b, and the depth of the water table is estimated to be approximately 21 m below ground surface \citep{thornley2019situ}. We note that the records from the sensor located at 18.3 m are contaminated with electric noise and consequently, were not considered in the analysis of this study.

\subsection{Ground Motion Information}
Acceleration time series were obtained from recorded ground motions with triaxial accelerometers at the surface and throughout the soil profile into the bedrock, as indicated by the black triangle symbols in \Cref{fig:Velocity_Profiles}. We select a suite of 32 and 52 low amplitude ground motions (both horizontal components from 16 and 26 unique events) at sites TIDA and DPDA, respectively. Ground motion records were selected to not exceed a peak ground acceleration (PGA) of 0.05 g. By including only low amplitude motions, we ensure that the challenges associated with modeling nonlinear soil behavior do not interfere with small-strain $V_s$, $V_p$, and damping estimation. These ground motions were selected from a larger candidate set based on a signal-to-noise ratio (SNR) criterion of greater than 3 dB between 0.5--10 Hz. This frequency range encompasses those frequencies that are of most interest from an engineering  perspective. Additionally, the expected  fundamental mode resonance frequencies at TIDA and DPDA are greater than the minimum screening frequency of 0.5 Hz and less than the maximum screening frequency of 10 Hz \citep{Hallal2021HVPart2}.
The acceleration time histories were filtered using a fourth-order zero-phase Butterworth filter with a passband of 0.5--10 Hz, which is the frequency range for which SNR $>$ 3 dB for all records.

In addition to using these processed ground motions to invert for subsurface material properties, the observed site response at each downhole array site was obtained using recordings from the deepest and the surface accelerometers. The site response is represented in this study using the ETF calculated as the ratio of the Fourier amplitude spectra (FAS) of the horizontal surface accelerations divided by the horizontal rock accelerations. The FAS were smoothed using a log-scale rectangular window, as described by \cite{goulet2014peer}, and the ETF for a single ground motion was then computed as the ratio of the smoothed FAS. The representative ETF for the site is finally taken as the geometric mean of all individual ETFs from the two horizontal components of all earthquake events. For a lognormal distribution, which is commonly assumed for the population of ETFs \citep{thompson2012taxonomy}, the geometric mean is equivalent to the median of the sample. We also compute the natural logarithmic standard deviation of the ETFs $(\sigma_{lnETF})$ at each frequency as a means to capture the variability in the ETFs. The ETFs at each site are presented later in the paper.

To evaluate the accuracy of the inverted subsurface material properties in modeling the site response, theoretical transfer functions (TTFs) are computed and compared with the median ETFs. Linear-viscoelastic, theoretical shear wave transfer functions were computed for the inverted $V_s$ profiles and damping ratios using the closed-form solution for a multi-layered damped soil over elastic bedrock \citep{Kramer_1996_Book}. The required input soil properties include $V_s$, mass density, and damping ratio for each layer. It should be noted that since the ETFs represent the surface-to-rock ratio in which the rock time histories include both upgoing and downgoing waves, the TTF is modeled to include those effects using what is referred to as the within boundary condition at the rock accelerometer \citep{Hallal2021HVPart2}.

\subsection{Dispersion Information} 
\begin{figure}[htbp!]
\centering
  \subfigure[]{\includegraphics[width = 0.49\textwidth]{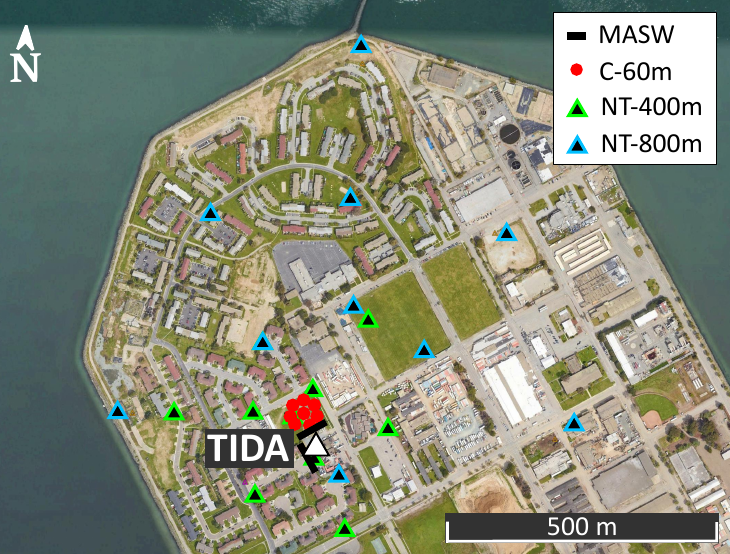}}
  \subfigure[]{\includegraphics[width = 0.49\textwidth]{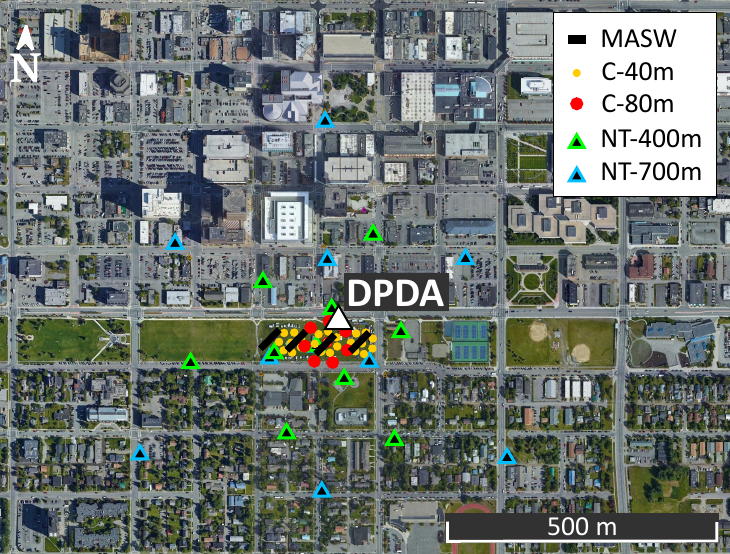}}
   \caption{Sensor arrays used to collect surface wave data in the vicinity of (a) TIDA and (b) DPDA. Testing was performed using active-source, linear-array Multi-channel Analysis of Surface Waves (MASW) and passive-source, circular (C) and nested triangular (NT) Microtremor Array Measurements (MAM). The numeric values in the legend of (a) and (b) correspond to the array apertures and individual symbols represent seismometer locations in each MAM array.}
   \label{fig:SW_Maps}
\end{figure}

Experimental dispersion data were obtained from surface-based noninvasive seismic measurements near the TIDA and the DPDA using active-source Multi-channel Analysis of Surface Waves (MASW) tests and several passive-source Microtremor Array Measurement (MAM) tests, as shown in \Cref{fig:SW_Maps}. We used identical equipment and followed similar procedures as those documented in \cite{teague2018measured} to extract experimental dispersion data across large areas at each site, and therefore limit the discussion on acquisition and processing herein. At TIDA, MASW testing involved two 46-m length linear arrays and MAM testing was performed using one circular and two nested triangular arrays of apertures ranging between 60 and 800 m (\Cref{fig:SW_Maps}a). At DPDA, MASW testing involved four 46-m length linear arrays and MAM testing was performed using four circular and two nested triangular arrays of apertures ranging between 40 and 700 m (\Cref{fig:SW_Maps}b). At each site, experimental Rayleigh wave dispersion data were computed separately for each individual array and were then averaged to develop representative mean experimental dispersion data with standard deviation estimates. The experimental dispersion data at each site (presented later in the paper) were of high quality and deemed to represent the fundamental Rayleigh mode.

\section{Synthetic Data Experiment}\label{sec:synthetic}
We first use synthetic data and simplified profiles to study the performance of the proposed methodology in estimation of $V_p$, $V_s$, and damping at the TIDA and DPDA sites; the application to real, measured data is presented in the next section. We note here that \cite{seylabi2020site} discussed the performance of the methodology for estimation of $V_s$ profile and damping ratio only. In this article, we extend this discussion to simultaneous estimation of $V_s$, $V_p$ and damping. As a result, we use dispersion data to inform $V_s$ and $V_p$ profile estimation and acceleration time series data to inform $V_s$ profile and damping estimation. We also discuss the importance of incorporating prior information of the water table depth on improving $V_p$ (and $V_s$) profile(s) and damping estimation.

To generate synthetic input data for the inversion (i.e., dispersion estimates and acceleration time series), we use a 1D subsurface model defined by the simplified $V_s$ and $V_p$ profiles shown in \Cref{fig:Velocity_Profiles}, and a constant damping ratio of $\xi=4$\% and density of 2000 kg/m$^3$ in each layer. The constant damping is a simplification, and as discussed previously, the forward problem is also based on an assumed depth-independent damping. First, we use Geopsy to compute the synthetic dispersion data. Second, we use one of the recorded acceleration time series at the bedrock (i.e., at a depth of 122 m for TIDA and 61 m for DPDA) to perform one dimensional site response analysis and calculate the synthetic acceleration time series at different strong motion sensor locations. To make the dispersion and acceleration time series data noisy, we perturb them with Gaussian noise with $\beta_1=0.03$ and $\beta_2=0.02$, respectively.
During the estimation, on the other hand, we use $\beta_1=0.05$ and $\beta_2=0.01$ to define the noise covariance $\Gamma$. We note here that, in real data, we expect the noise due to measurement and modeling errors to be higher in acceleration time series compared to dispersion data, and this is why we set $\beta_1 > \beta_2$.

\subsection{TIDA Site} 
We try to retrieve the assumed 1D subsurface model  at the TIDA site by joint inversion of the synthetic dispersion and acceleration time series that were calculated from the assumed model. We discretize the soil column with $l=27$ soil layers with thickness ranging from 2 to 7 m. The considered discretization is based on refining the near surface layer with 2-m thickness, as obtained from one-third of the minimum wavelength considered in the synthetic dispersion data \citep{cox2016layering}, and then using layers with constant thicknesses of 5 m, except where a layer interface was needed based on the expected water table or bottom depth. Furthermore, for the problem at hand, as discussed in detail in \cite{seylabi2020site}, the ensemble Kalman inversion results are generally independent of the number of layers one uses for discretizing the domain and a coarser model follows the same trend as the refined one. This insensitivity is partially due to the imposed constraints that improves the inverse problem well-posedness. As shown in \Cref{fig:Velocity_Profiles}, the target $V_s$ and $V_p$ profiles do not necessarily increase with depth. Nonetheless, the velocity reversals are not significant, and therefore, for inverse analysis at this site, we decide to enforce the following constraints to keep both profiles positive and monotonically increasing with depth. This helps the algorithm to search for optimal solutions in a smaller space of parameters, and therefore, improve the inverse problem well-posedness.
\begin{align}
&V_{s,1} \geq100\text{~m/s}, \quad V_{s,l} \leq 3000\text{~m/s}\\
&V_{s,i}\leq V_{s,i+1} \quad i = 1,\dots,l-1\\
&V_{p,i}\leq V_{p,i+1} \quad i = 1,\dots, l-1\\
&0.1\% \leq \xi \leq 20\%\,.
\end{align}  
Furthermore, to study the effects of the water table depth, we consider the following two cases of (1) not including prior information on water table, and (2) including prior information on water table, as follows:
\begin{align}
\text{Case 1} \rightarrow \quad 0\leq z \leq 122: \quad V_{p} \geq 1.6 V_{s},\quad
\text{Case 2}\rightarrow \quad 
\begin{cases}
0\leq z \leq 6: & V_p\geq1.6V_s\\
6\leq z \leq 85: & V_p \geq 5.0 V_s\\
85\leq z\leq 122: & V_p \geq 1.6 V_s
\end{cases}
\end{align}
where coefficients 1.6 and 5.0 set the minimum Poisson's ratio equal to 0.18 and 0.48, respectively, with the latter being consistent with typical values for soil below the water table. For each case, we use the noisy synthetic data to estimate the $V_s$ and $V_p$ profiles and damping considering the enforced constraints.

\begin{figure}[htbp!]
\centering
  \subfigure{\includegraphics[width = 0.49\textwidth]{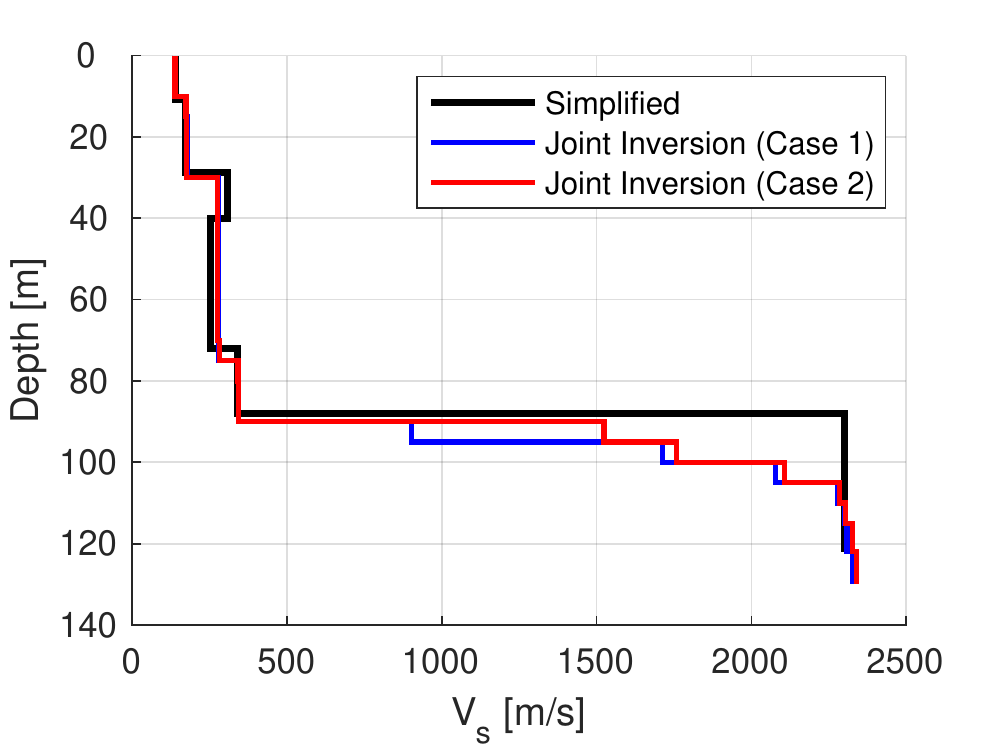}}
  \subfigure{\includegraphics[width = 0.49\textwidth]{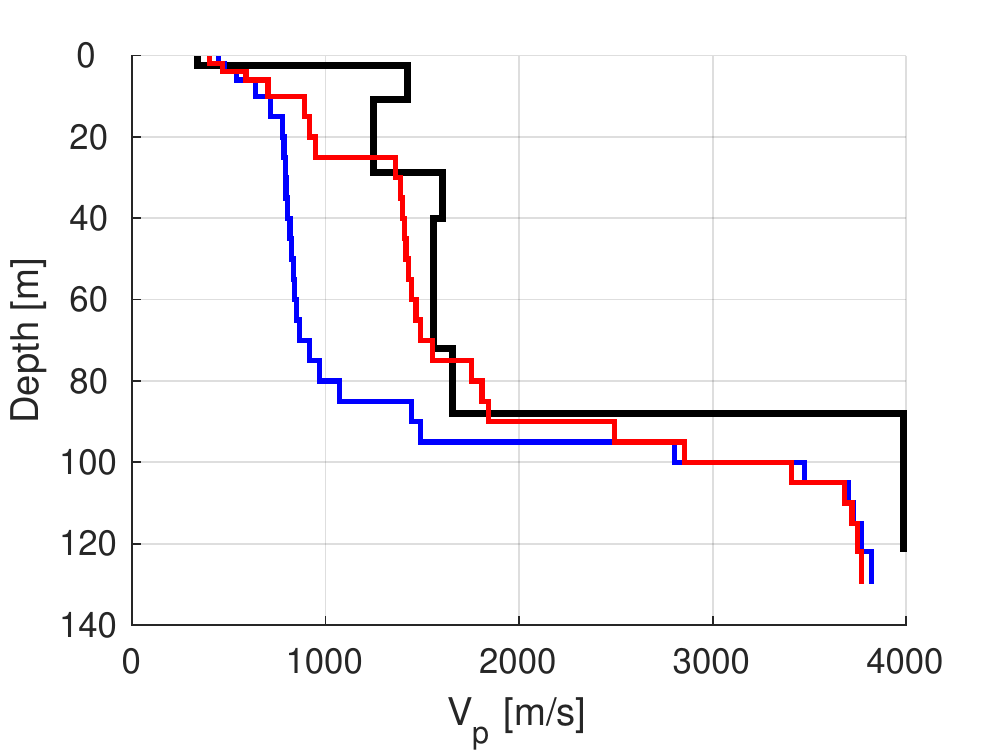}}
   \caption{Estimated $V_s$ and $V_p$ profiles at TIDA site using synthetic data.}
   \label{fig:TIDA-syn-final}
\end{figure}

\begin{figure}[htbp]
\centering
  \subfigure{\includegraphics[width = 0.49\textwidth]{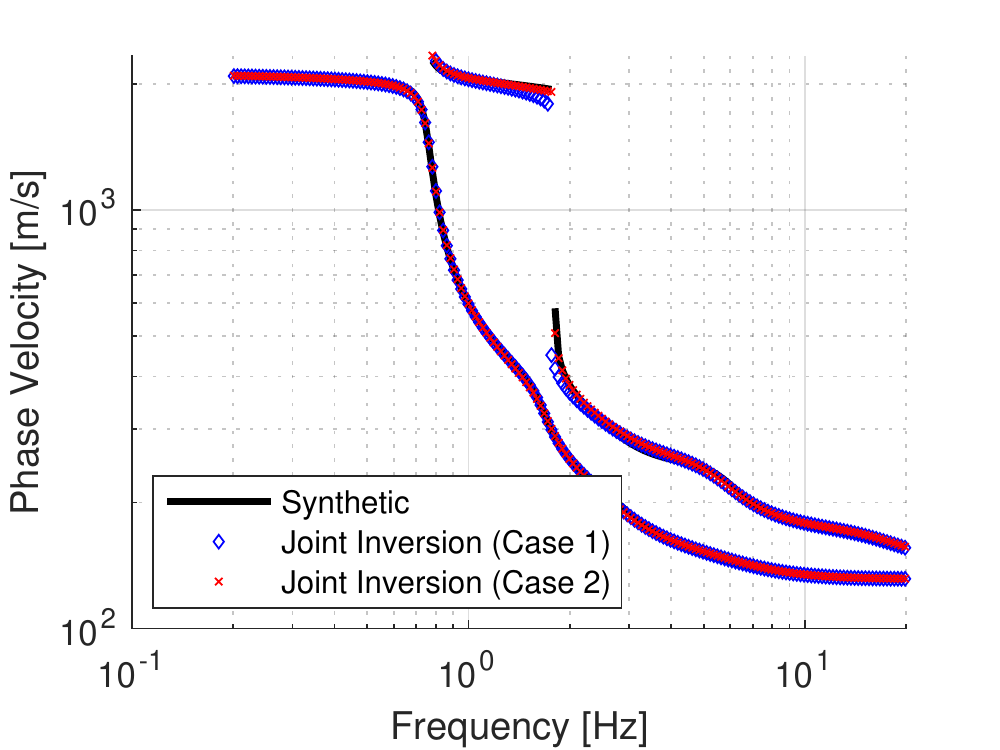}}
  \subfigure{\includegraphics[width = 0.49\textwidth]{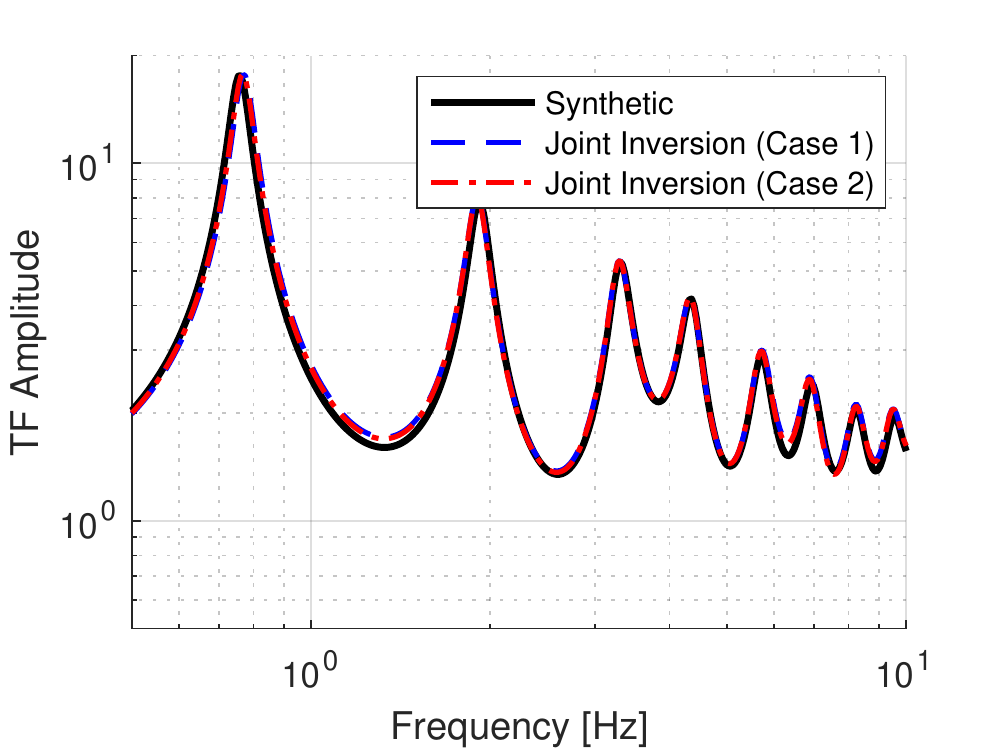}}
   \subfigure{\includegraphics[trim = 2cm 0 1.5cm 0,clip,width = \textwidth]{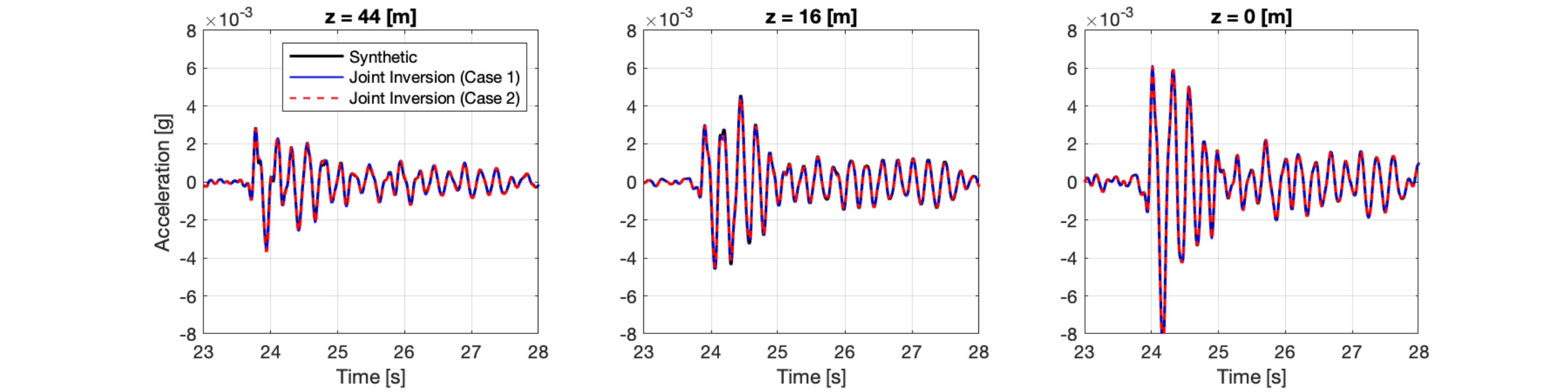}}
   \caption{Performance of estimated model in predicting synthetically generated dispersion data, site transfer function, and acceleration time-series at TIDA site.}
   \label{fig:TIDA-syn-predict}
\end{figure}

\Cref{fig:TIDA-syn-final} shows the estimated $V_s$ and $V_p$ profiles for each case in comparison with the simplified/assumed subsurface model, and \Cref{fig:TIDA-syn-predict} shows the performance of these two cases in capturing the synthetic data without noise. As shown, although the results from Case 1 and Case 2 are different in terms of $V_p$, these differences have minimal effects on the fundamental mode of the computed dispersion curve and the acceleration time series. However, the estimation results suggest that enforcing a priori knowledge about the water table depth can help improve the estimation of the $V_p$ profile. By comparing the phase velocity of the first higher Rayleigh mode, we notice that, although this mode was not included in the inverse analysis, it is well captured by the results of Case 2. However, the results of Case 1 are less successful in capturing the higher Rayleigh mode at frequencies between 1--2 Hz. These results suggest that incorporating higher modes and water table information can help improve the estimation of $V_p$ profiles.
We also note that the final ensemble means of the estimated depth-independent damping for Case 1 and Case 2 are 4.04\% and 4.02\%, respectively, with the target value used in the synthetic data generation being 4\%.

\subsection{DPDA Site} 
As a second experiment, we try to retrieve the assumed 1D subsurface model at the DPDA site using synthetic acceleration time series and dispersion data. We used simplified $V_p$ and $V_s$ profiles of \Cref{fig:Velocity_Profiles} in combination with an assumed 4\% damping ratio for all layers to generate the synthetic data. As shown, the target $V_s$ and $V_p$ profiles are not necessarily increasing with depth and the velocity reversals are more significant compared to TIDA. Therefore, for this site, we decided to relax the enforced constraints to allow the $V_s$ and $V_p$ at layer $i$ to be at most 50\% higher than the values at layer $i+1$. That is:
\begin{align}
&V_{s,i}\leq 1.5V_{s,i+1} \quad i = 1, \dots, l-1\\
&V_{p,i}\leq 1.5V_{p,i+1} \quad i = 1,\dots, l-1\\
&V_{s,1}\geq 100, \quad V_{s,l} \leq 2000\\
&0.1\% \leq \xi \leq 20\%\,.
\end{align}
We note that $l=31$ is the total number of considered layers with thickness of 2 m, except the last layer which is 3 m. Again, to study the effects of the water table depth, we consider the following two cases:
\begin{align}
\text{Case 1} \rightarrow \quad 0\leq z \leq 61: \quad V_{p} \geq 1.6 V_{s},\quad
\text{Case 2}\rightarrow \quad 
\begin{cases}
0\leq z \leq 20: & V_p\geq1.6V_s\\
20\leq z \leq 40: & V_p \geq 5.0 V_s\\
40\leq z\leq 61: & V_p \geq 1.6 V_s
\end{cases}
\end{align}

\begin{figure}[htbp!]
\centering
  \subfigure{\includegraphics[width = 0.49\textwidth]{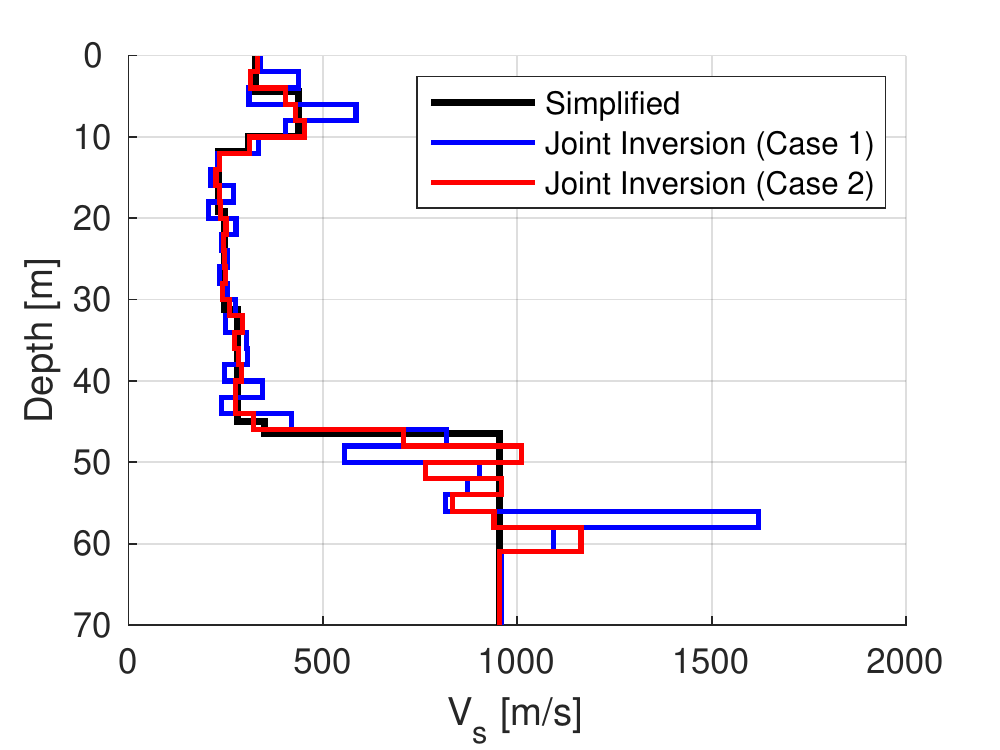}}
  \subfigure{\includegraphics[width = 0.49\textwidth]{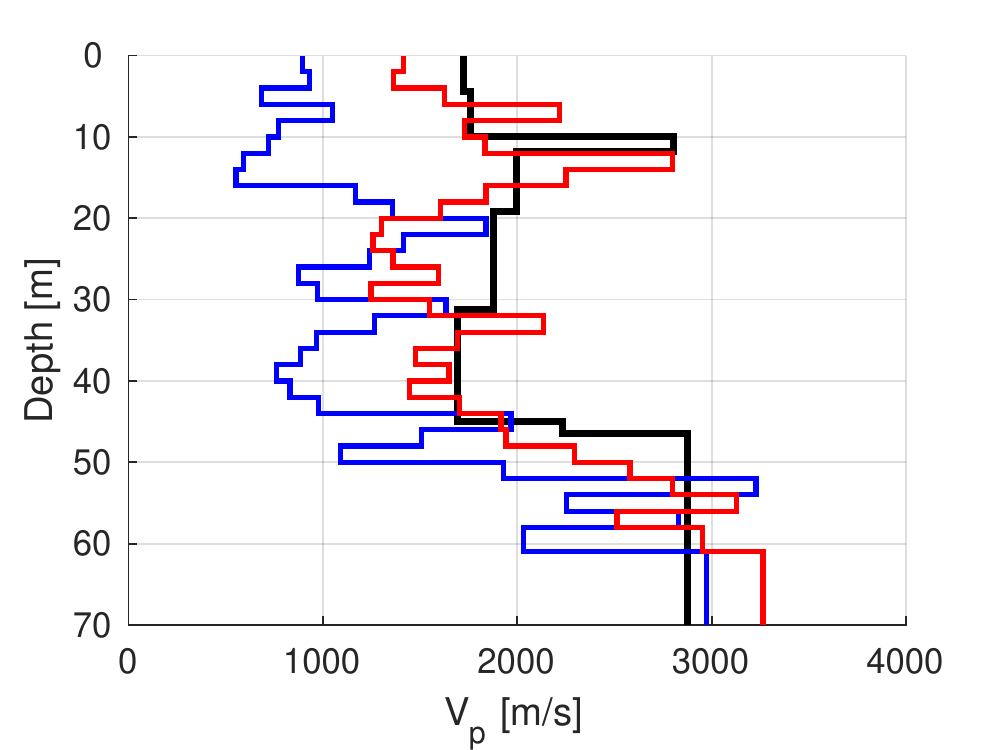}}
   \caption{Estimated $V_s$ and $V_p$ profiles at DPDA site using synthetic data.}
   \label{fig:DPDA-syn-final}
\end{figure}

\begin{figure}[htbp]
\centering
  \subfigure{\includegraphics[width = 0.49\textwidth]{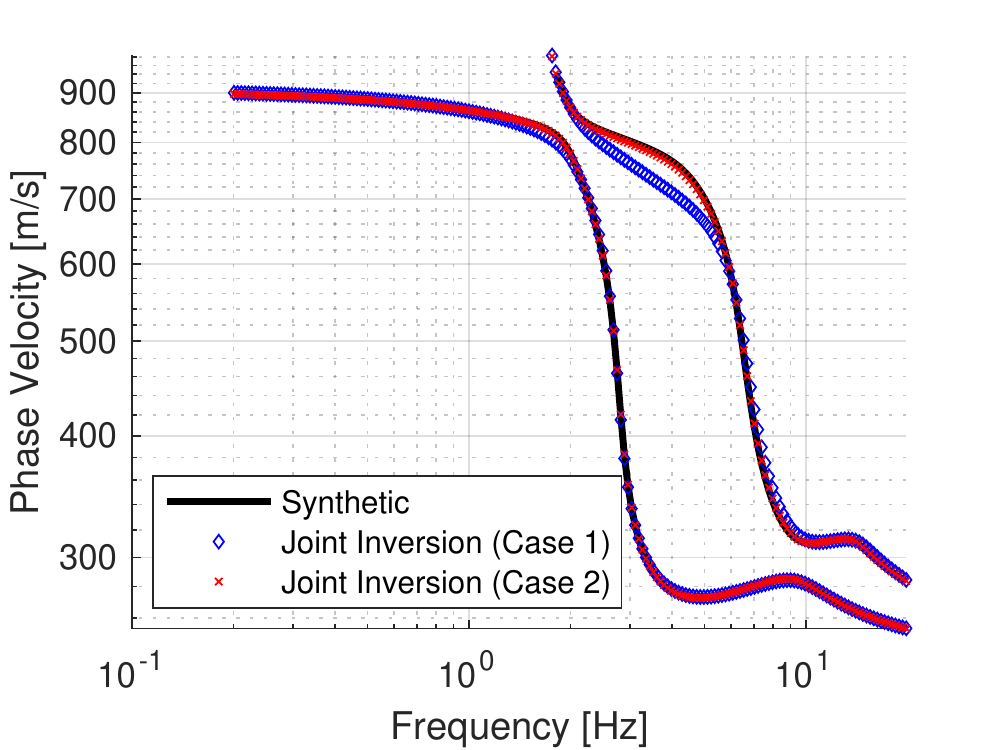}}
  \subfigure{\includegraphics[width = 0.49\textwidth]{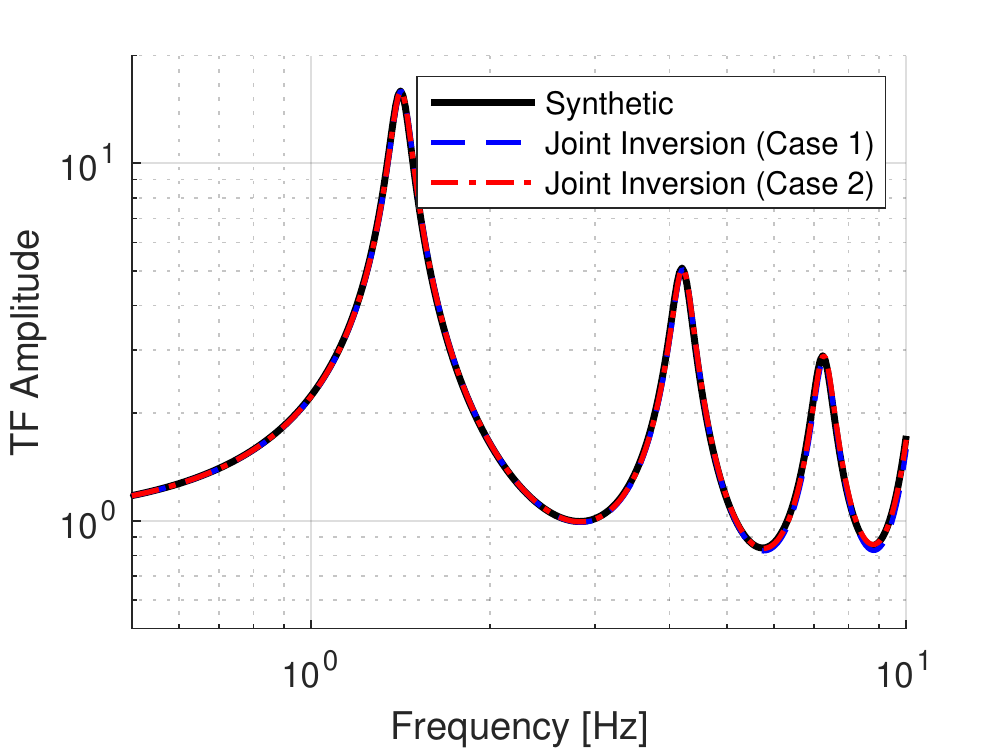}}
   \subfigure{\includegraphics[trim = 2cm 0 1.5cm 0,clip,width = \textwidth]{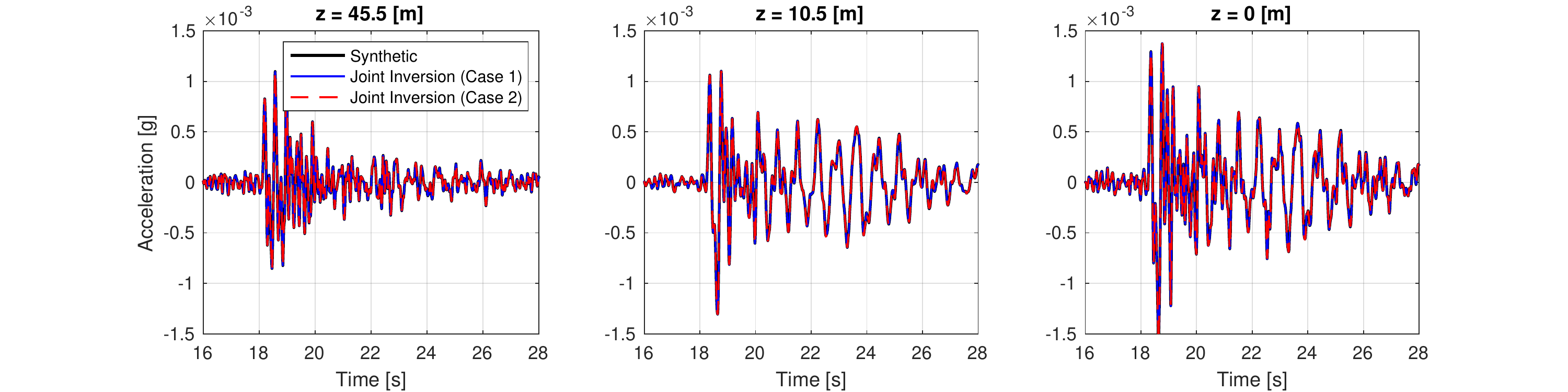}}
   \caption{Performance of estimated model in predicting synthetically generated dispersion data, site transfer function, and acceleration time-series at DPDA site.}
   \label{fig:DPDA-predict}
\end{figure}

The estimated $V_s$ and $V_p$ profiles are shown in \Cref{fig:DPDA-syn-final} relative to the simplified/assumed subsurface model and the performance of these inversion cases in capturing the synthetic acceleration time histories and dispersion data is presented in \Cref{fig:DPDA-predict}. As shown, the estimated $V_s$ profiles agree well with the target profile and the estimated $V_p$ profiles reasonably capture the target $V_p$ profile considering the enforced constraints, particularly for Case 2. Similar to the results at TIDA, it appears that including higher mode dispersion data can help improve the estimation of the $V_p$ profile, as evident in the better match between the synthetic data and those for Case 2 between 2--5 Hz (\Cref{fig:DPDA-predict}). The estimated damping using synthetic data is 3.9\% for both cases and in very good agreement with the target value of 4\%.

Both synthetic experiments at TIDA and DPDA suggest that the proposed algorithm can successfully estimate the $V_s$ profile and damping ratio even for more complicated subsurface conditions, where the $V_s$ and $V_p$ profiles do not always increase with depth. Also, as discussed, the enforced constraints and a priori knowledge on the water table depth can help improve the $V_p$ estimation, which can in turn improve the $V_s$ estimation, as the forward problems are dependent on both $V_p$ and $V_s$.

\section{Real Data Experiment}\label{sec:gvda}
In this section, we present the results of characterizing the 1D subsurface model at TIDA and DPDA using the proposed methodology and real data.

\subsection{TIDA Site}
To estimate the $V_p$, $V_s$, and damping ratio at the TIDA site, we follow the same strategy as in the synthetic data experiment, and consider two cases to study the effects of incorporating water table depth information. For Case 1, we consider discretizing the domain with 25 layers with thickness of 5 m, and for Case 2, we use 27 layers with thicknesses between 2 and 7 m, similar to the synthetic experiment. Therefore, in Case 1 we estimate 51 parameters and for Case 2 we estimate 55 parameters. In total, for each case, we perform joint inversion for each earthquake event and for both horizontal components (i.e., 32 inverse analyses). Furthermore, to study the effects of using complementary data, we performed inverse analysis using dispersion data only (i.e., without acceleration time series).

\begin{figure}[htbp!]
\centering
  \subfigure[Case 1]{\includegraphics[width = 0.49\textwidth]{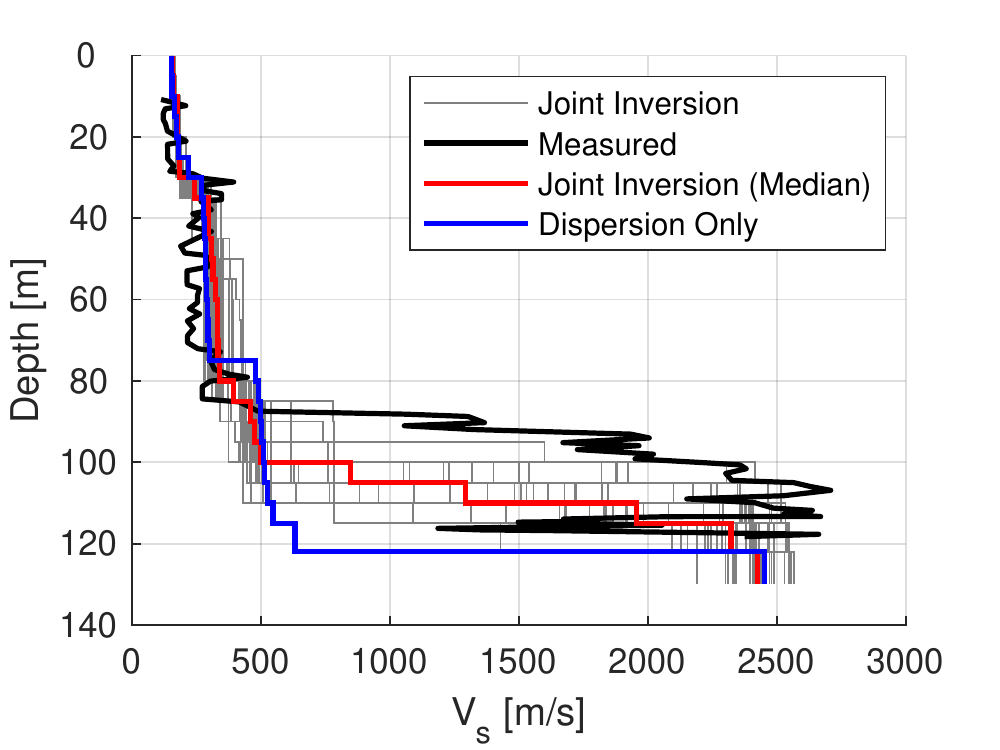}}
  \subfigure[Case 1]{\includegraphics[width = 0.49\textwidth]{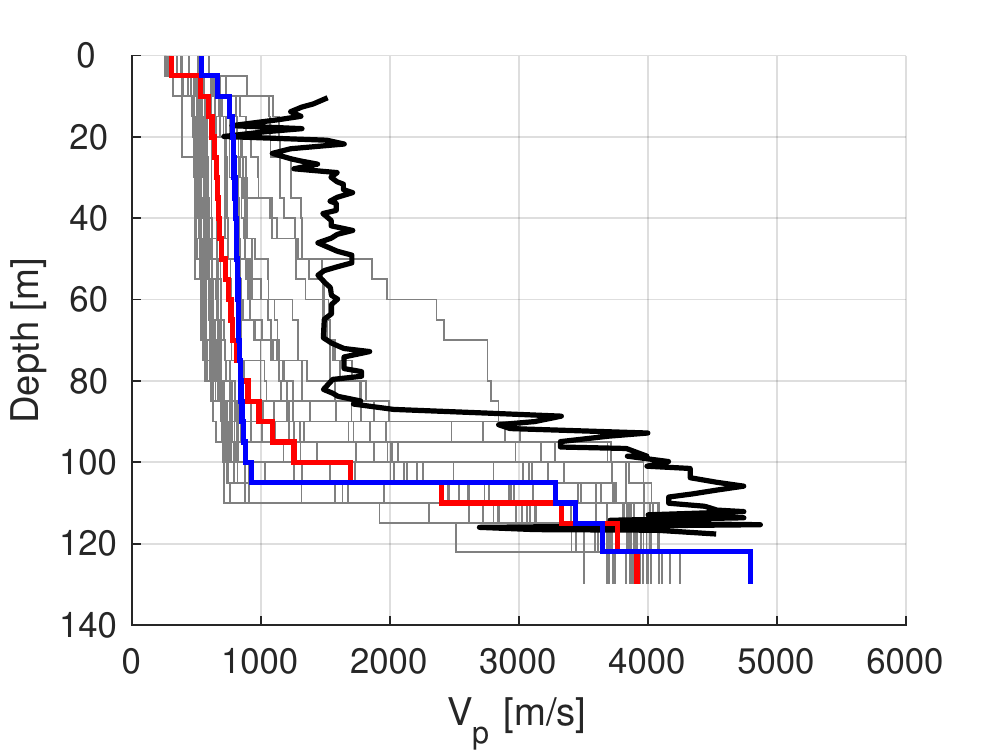}}
  \subfigure[Case 2]{\includegraphics[width = 0.49\textwidth]{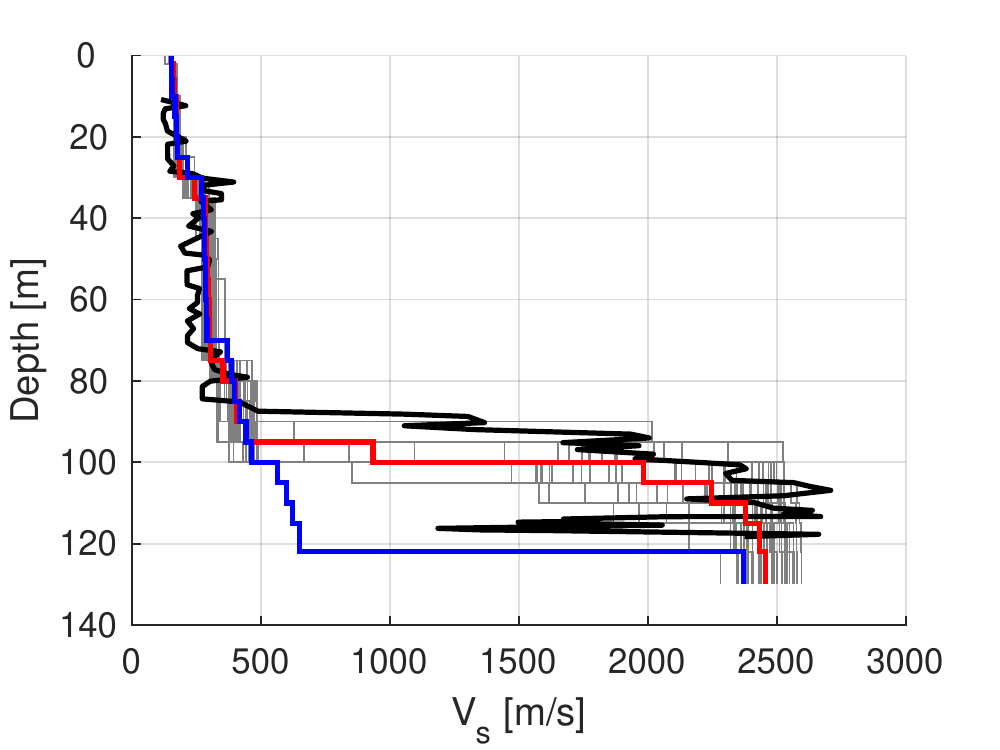}}
  \subfigure[Case 2]{\includegraphics[width = 0.49\textwidth]{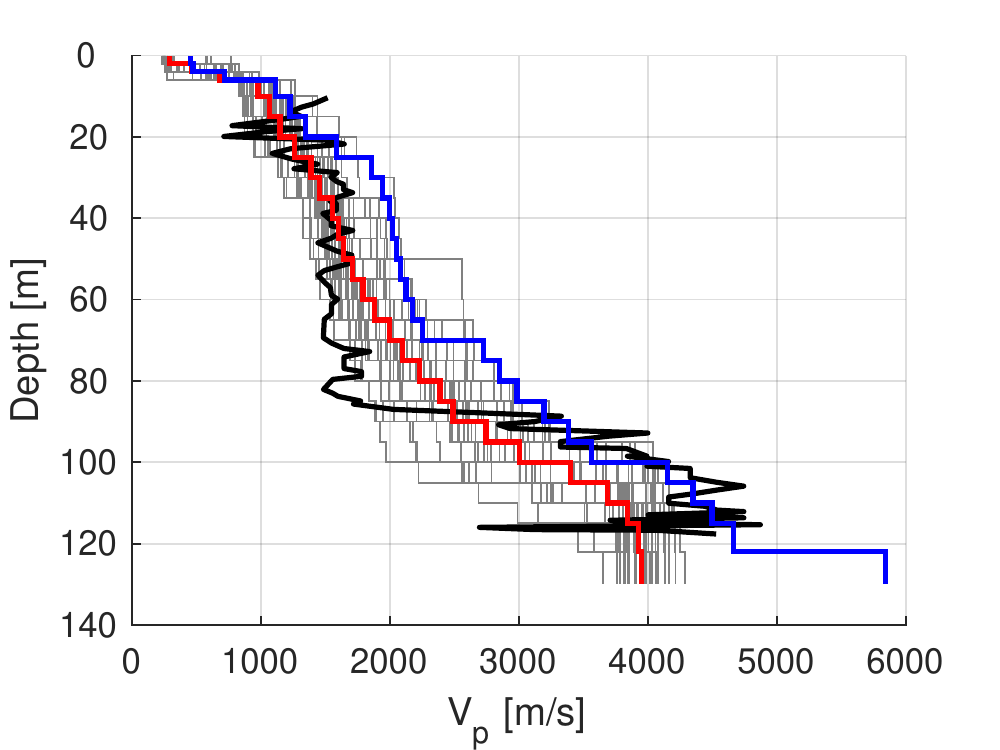}}
   \caption{$V_s$ and $V_p$ estimation at TIDA site and for Case 1 and Case 2.}
   \label{fig:TIDA-real-vsvp}
\end{figure}

\Cref{fig:TIDA-real-vsvp} shows the median of final ensembles of the estimated $V_s$ and $V_p$ profiles compared to the measured profiles at the site. To compute the median $V_s$ and $V_p$ profiles from all 32 inversions, we assume a lognormal distribution. As shown in \Cref{fig:TIDA-real-vsvp}, inverting dispersion data alone using the proposed algorithm is not successful at estimating the depth of the sharp velocity contrast at a depth of 80--90 m, as indicated by the abrupt increase in $V_s$, regardless of whether prior information on water table depth is included or not (i.e., Case 1 or 2). It should be noted that these dispersion-only inversion results, particularly the depth to bedrock, are in poorer agreement with the measured $V_s$ profile than those obtained by \cite{hallal2021comparison}, wherein they obtained $V_s$ profiles by joint inversion of dispersion and HVSR data that agreed very well with the depth to bedrock and measured $V_s$ profile at TIDA. Similarly herein, joint inversion of the dispersion data and acceleration time series with the proposed algorithm is more successful in capturing the sharp velocity contrast observed in the PS logging and including prior information on water table (i.e., Case 2) significantly improves both the $V_p$ and $V_s$ estimation.

\subsection{DPDA Site}
\begin{figure}[htbp!]
\centering
  \subfigure[Case 1]{\includegraphics[width = 0.49\textwidth]{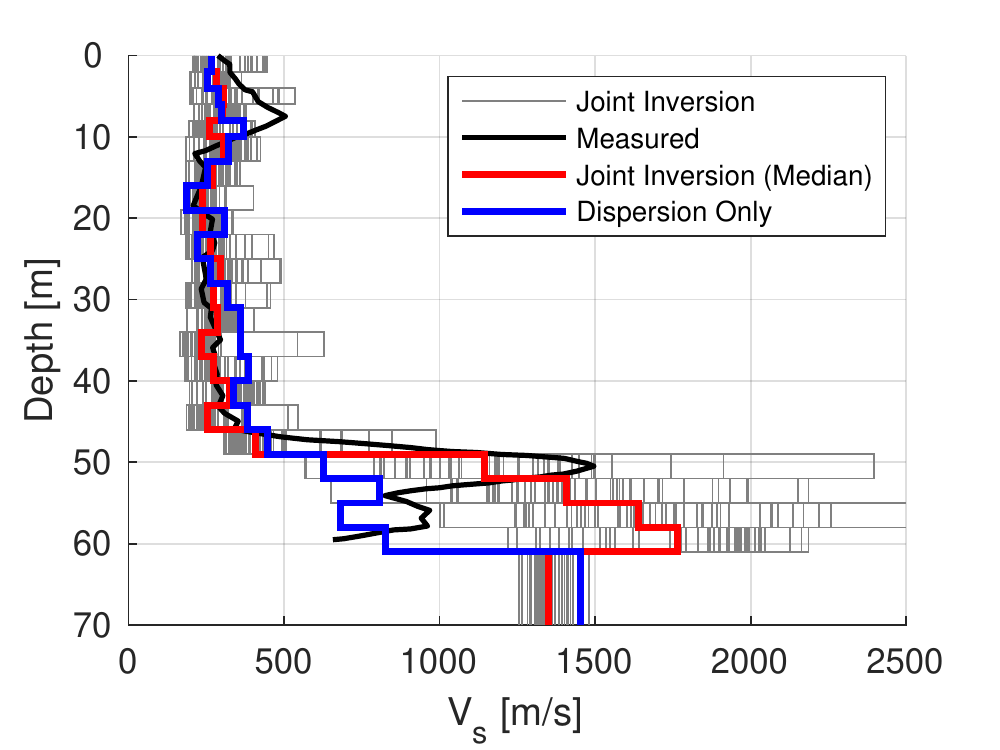}}
  \subfigure[Case 1]{\includegraphics[width = 0.49\textwidth]{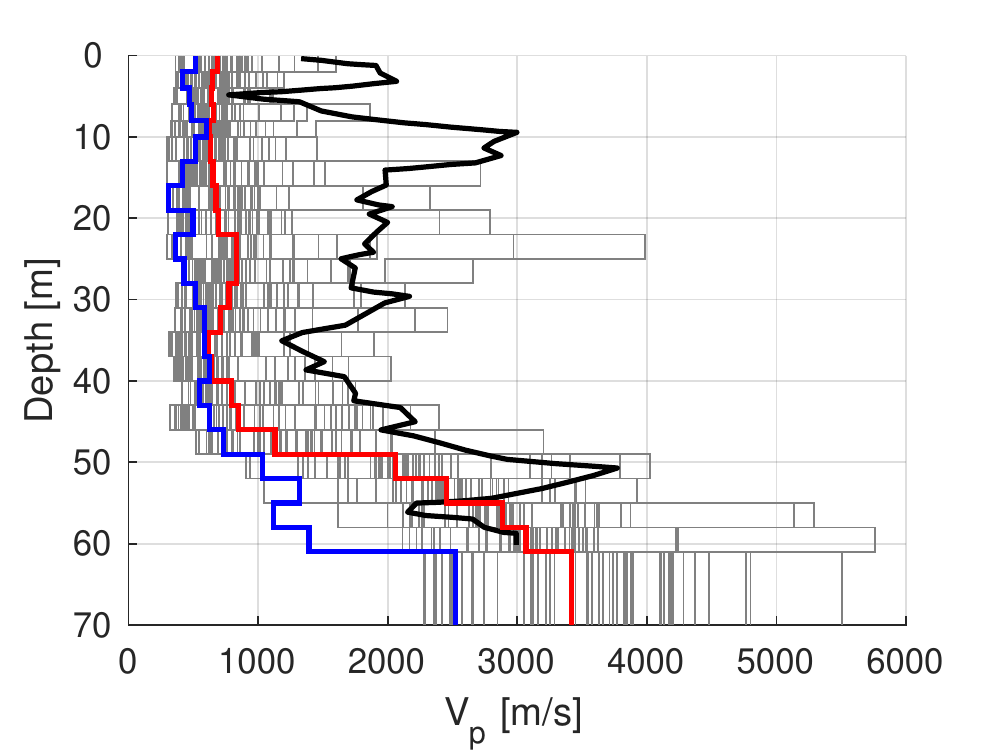}}
  \subfigure[Case 2]{\includegraphics[width = 0.49\textwidth]{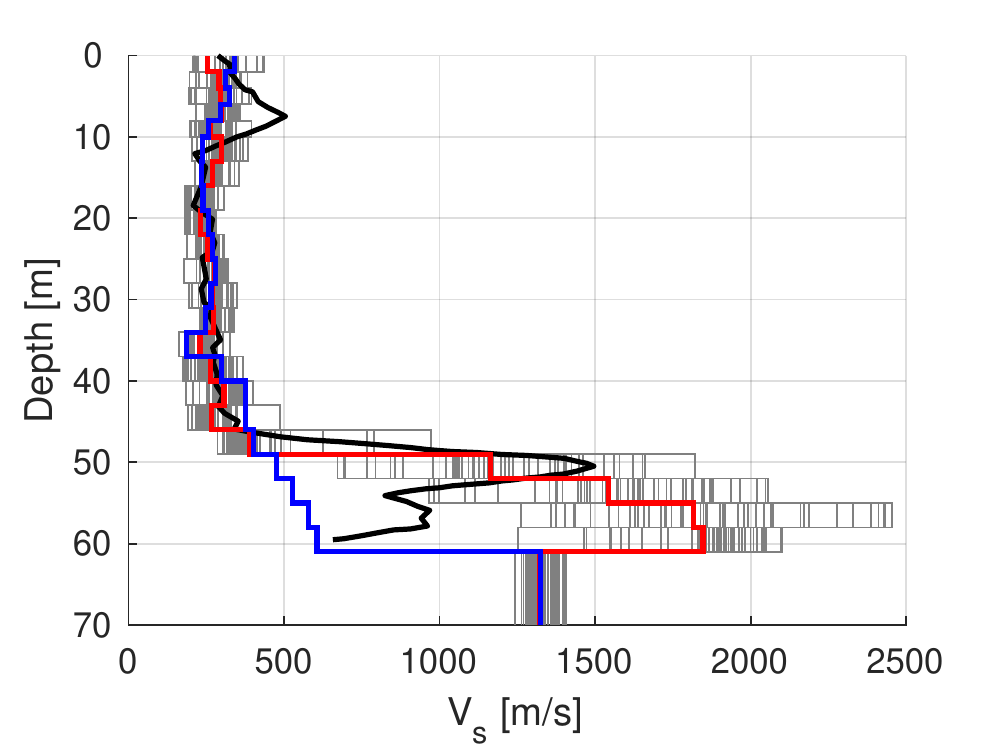}}
  \subfigure[Case 2]{\includegraphics[width = 0.49\textwidth]{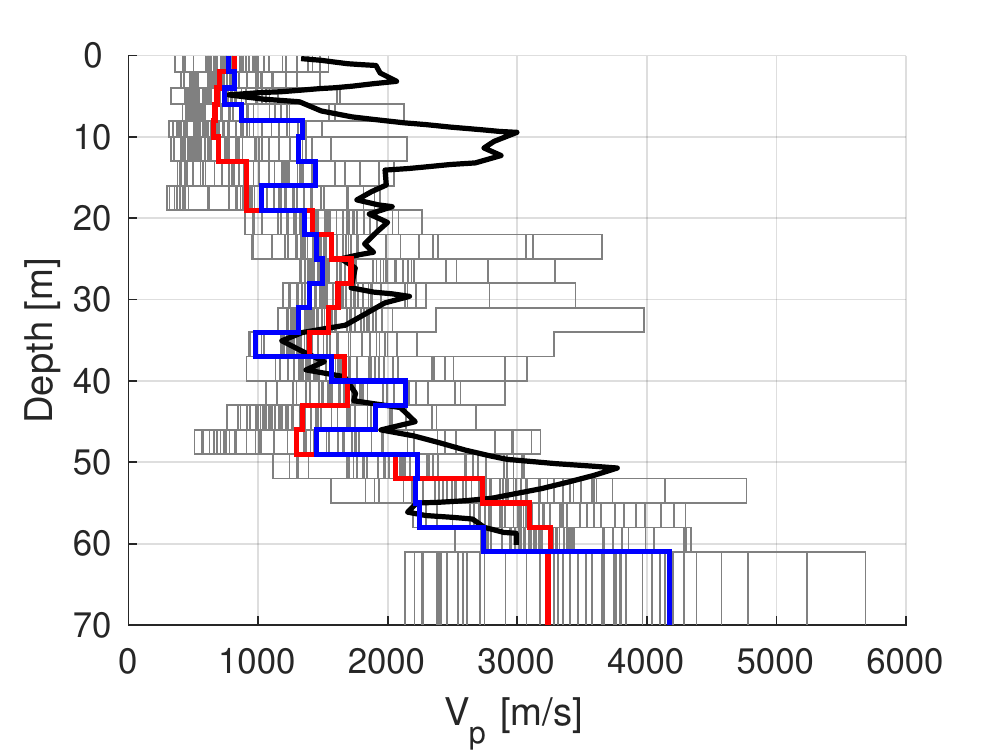}}
   \caption{$V_s$ and $V_p$ estimation at DPDA site and for Case 1 and Case 2.}
   \label{fig:DPDA-real-vsvp}
\end{figure}

To estimate the $V_p$, $V_s$, and damping ratio at the DPDA site, we discretized the soil deposit with 23 layers, similar to the synthetic data experiment, with the same set of enforced constraints. \Cref{fig:DPDA-real-vsvp} shows the median of estimated $V_s$ and $V_p$ profiles compared to available seismic downhole measurements. Again, to compute the median $V_s$ and $V_p$ profiles from all joint inversion results we assume a lognormal distribution. As shown in \Cref{fig:DPDA-real-vsvp}, inverting dispersion data alone using the proposed algorithm is not as successful at estimating the velocity contrast at the depth of 40--50 m, which is similar to the results reported at TIDA. However, similar to what was discussed above in regards to TIDA, we note that \cite{hallal2021comparison} achieved a good match to the depth of the velocity contrast at DPDA by inverting dispersion and HVSR data. Also, similar to TIDA, we notice that by enforcing the water table constraint (i.e., Case 2), $V_p$ estimation improves significantly. However, in contrast to the results from TIDA, Case 2 does not result in a significant change to the $V_s$ profile. 

\section{Discussion of Results}\label{sec:discuss}

\subsection{Assessment Against Experimental Data}

In this section, we assess the performance of the inversion results against the experimental Rayleigh wave dispersion data (obtained from the arrays shown in \Cref{fig:SW_Maps}), the empirical transfer function between the surface and deepest accelerometers in the downhole array, and the recorded ground motions at different depths in the downhole array. For brevity, we only present results from Case 2 for both joint inversion and dispersion data inversion only.

\begin{figure}[htbp!]
\centering
  \subfigure[]{\includegraphics[width = 0.49\textwidth]{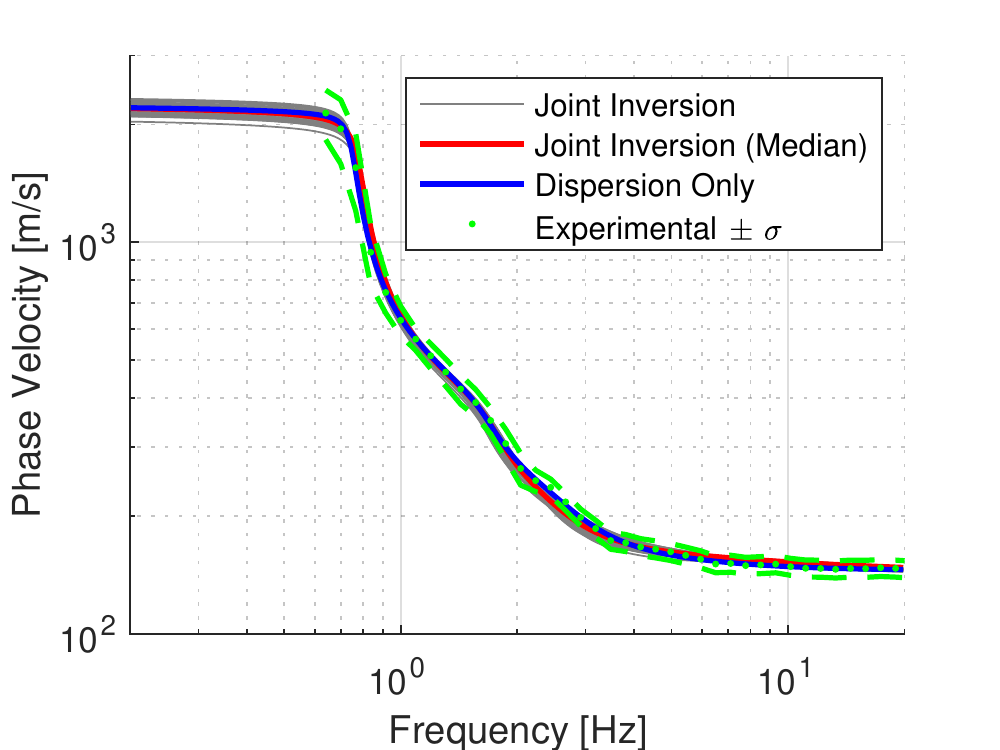}}
  \subfigure[]{\includegraphics[width = 0.49\textwidth]{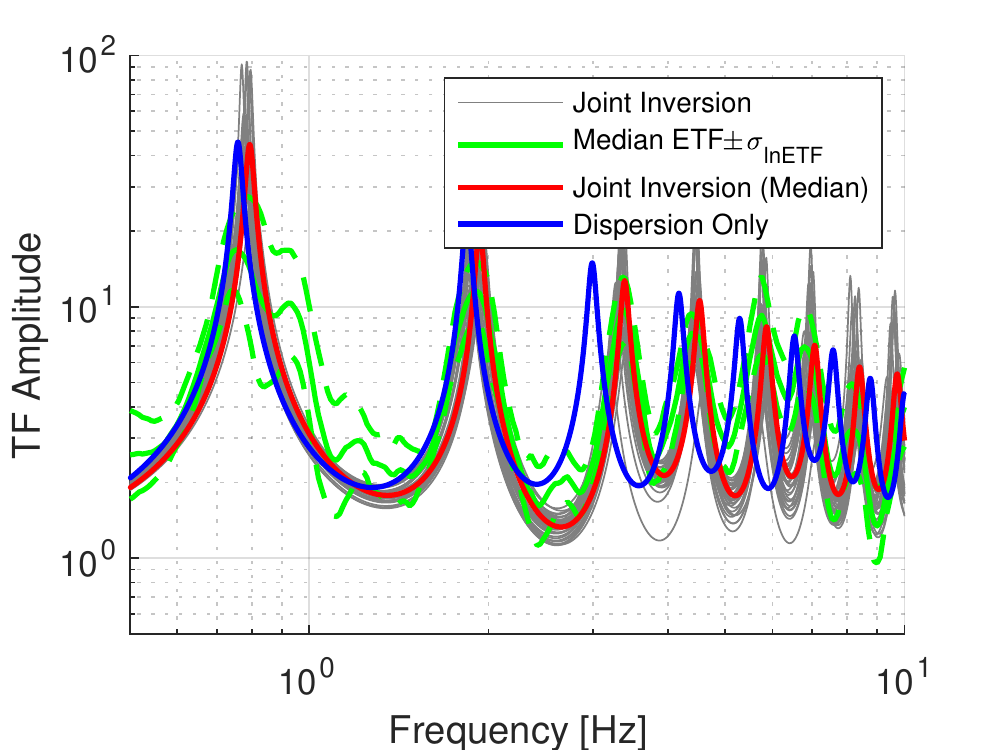}}
   \subfigure[]{\includegraphics[trim = 3.5cm 0cm 3.3cm 0cm, clip,width =\textwidth]{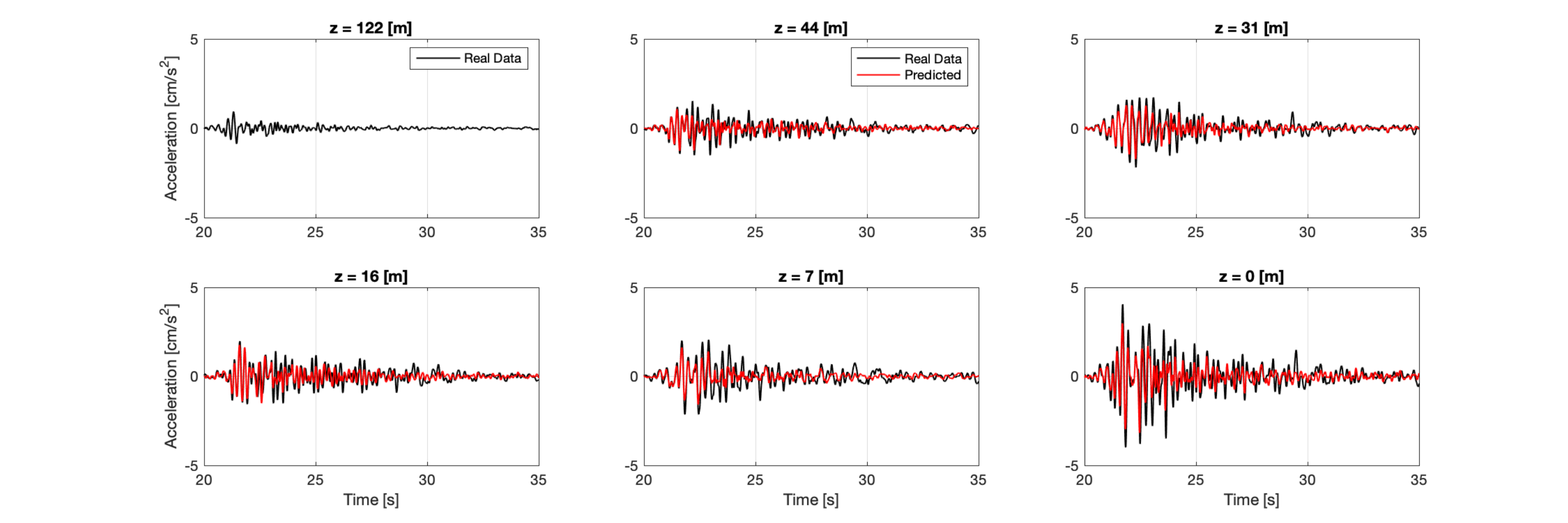}}
   \caption{Model performance in predicting the experimental dispersion data and the empirical transfer function at TIDA site.}
   \label{fig:TIDA-real-predict}
\end{figure}

\noindent
\underline{TIDA Site:}
\Cref{fig:TIDA-real-predict} shows the performance of the results from Case 2 in capturing the experimental dispersion data, ETFs and recorded ground motions. It is evident, qualitatively, that the estimated subsurface models by inversion of dispersion data only and joint inversion of dispersion and acceleration time series data are both in good agreement with the experimental data. To provide a quantitative measure of the agreement between the experimental dispersion data and the inversion results, a dispersion misfit value was computed for each inversion subsurface model, as described in \cite{wathelet2004surface}. The joint inversion results had misfit values ranging between 0.36--0.78, whereas the dispersion only inversion results had a misfit of 0.21. These dispersion misfit values support the good agreement with the experimental dispersion data, and indicate that, on average, the theoretical dispersion curves fall within less than one standard deviation from the mean trend of the experimental dispersion data.

Arguably, the TTFs of the joint inversion results match better with the median ETF than those from the dispersion only inversion results, particularly at higher modes (\Cref{fig:TIDA-real-predict}b). This result is expected, as the joint inversion results are based on the recorded acceleration time series in addition to the experimental dispersion data, and the more complimentary pieces of information used to constrain an inversion the better. A quantitative assessment of the goodness-of-fit between the TTFs and median ETF is presented later in the paper in the section titled \S Comparison with Other Approaches Used to Incorporate Wave Scattering into 1D GRAs.

The accuracy of the predicted site response using the subsurface properties obtained from joint inversion can also be evaluated by comparing acceleration time histories. \Cref{fig:TIDA-real-predict}c shows a comparison between the predicted and recorded time series at the different accelerometer depths for a sample ground motion at TIDA. Note that there is no predicted time history at the deepest accelerometer depth (z = 122 m), since this is used as the input ground motion for prediction, as discussed in \S Methodology. In addition, the predictions are based on the median $V_p$, $V_s$, and damping ratio obtained from joint inversion. It is visually evident that the predicted time series are in good overall agreement with those recorded at different depths. This further corroborates that the joint inversion algorithm used in this article is capable of reasonably reproducing the recorded acceleration time series when using real data.
 
\begin{figure}[htbp!]
\centering
  \subfigure[]{\includegraphics[width = 0.49\textwidth]{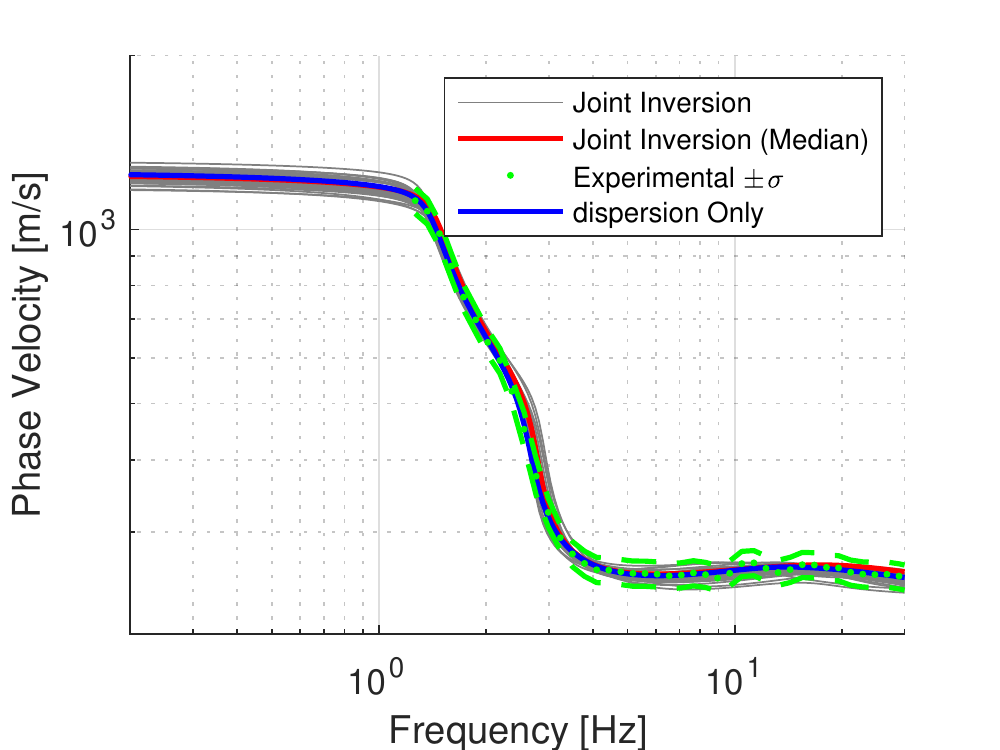}}
  \subfigure[]{\includegraphics[width = 0.49\textwidth]{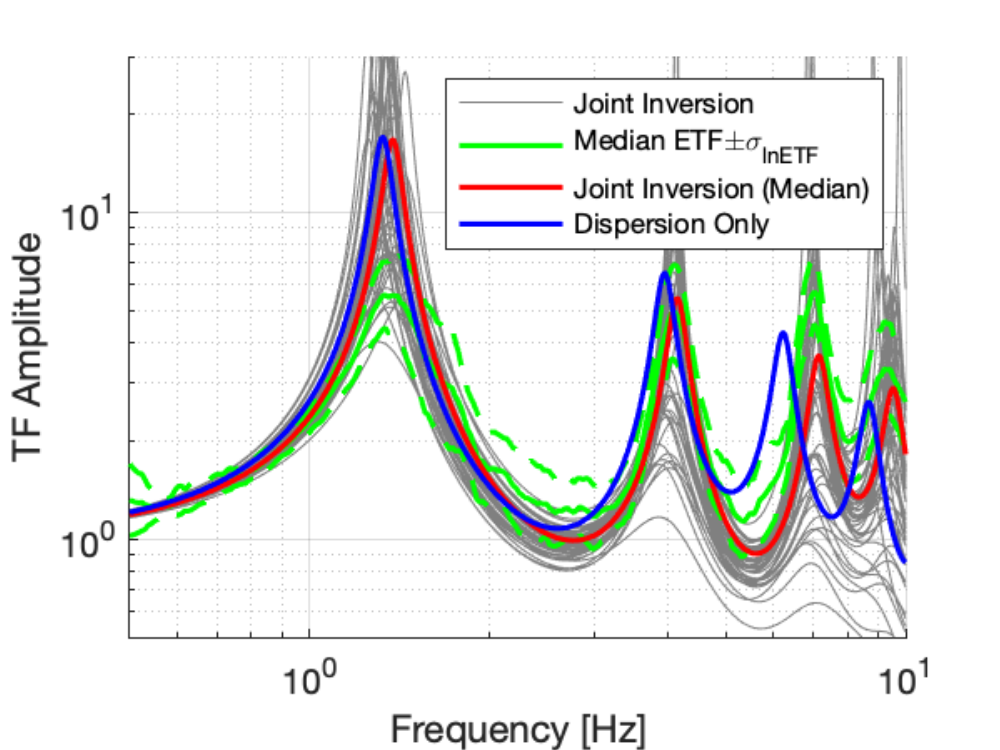}}
  \subfigure[]{\includegraphics[trim = 3.5cm 0cm 3.3cm 0cm, clip,width =\textwidth]{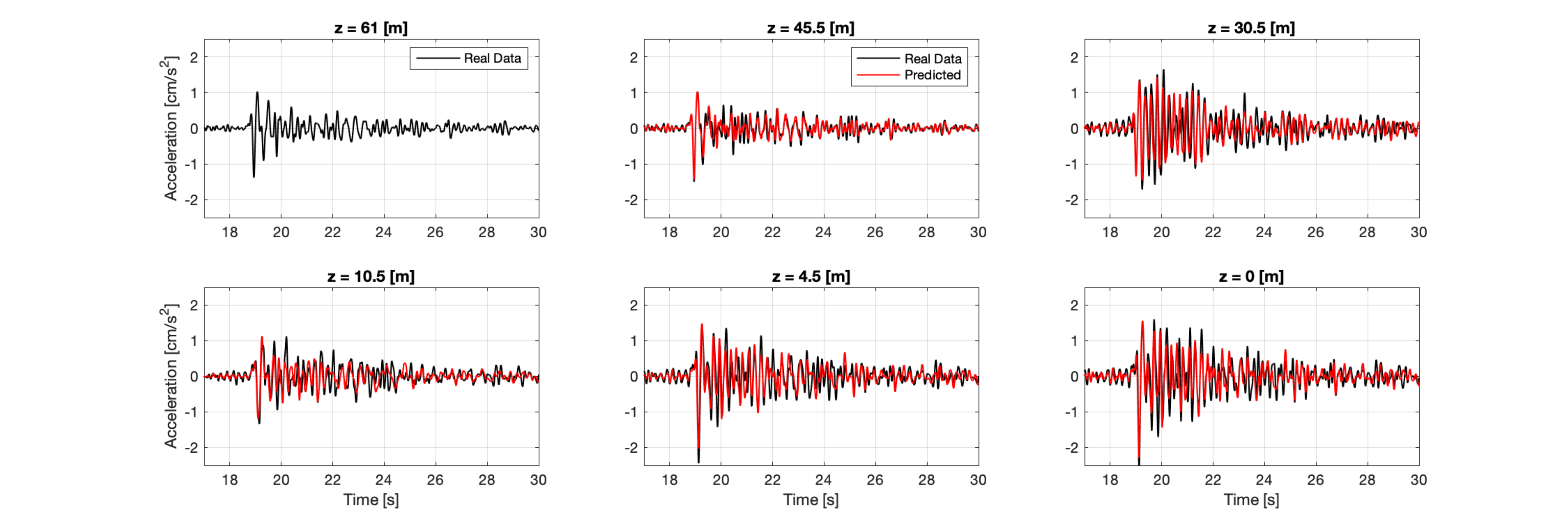}}
   \caption{Model performance in predicting the experimental dispersion data and the empirical transfer function at DPDA site.}
   \label{fig:DPDA-real-predict}
\end{figure}

\noindent
\underline{DPDA Site:}
\Cref{fig:DPDA-real-predict} shows the performance of the results from Case 2 in capturing the experimental Rayleigh wave dispersion data, the ETF from the recorded ground motions at  DPDA, and the recorded motion time series. Similar to  TIDA, dispersion misfit values were computed, and those from joint inversion had values between 0.28--1.25, whereas the estimated model from dispersion inversion had a misfit of 0.19. As noted above in regards to the results at TIDA, these low dispersion misfit values indicate a good overall agreement between the estimated models and the experimental dispersion data. It is also evident that the results from joint inversion are superior to those from dispersion only inversion in matching the median ETF, and clearly in the better agreement with the ETF's higher modes, consistent with the results at TIDA. The good fit of the site response predictions based on the joint inversion results are also evident in the sample time series comparison shown in \Cref{fig:DPDA-real-predict}c.

\subsection{Assessment of Estimated Damping Ratio}
A key advantage of the algorithm used in this article for characterizing the subsurface material is the ability to estimate the in situ damping. Several researchers have hypothesized that subsurface heterogeneity and three-dimensional scattering by discontinuous, sloping, or generally heterogeneous layers causes attenuation of the seismic waves and an apparent loss of energy. Lab-based damping measurements, however, only capture material damping in a small specimen and are incapable of capturing these complex mechanisms encountered in the field. For example, \citet{assimaki2007inverse}, in a study at five downhole array sites, observed that the back-calculated damping values are substantially higher than those estimated from lab measurements, which was attributed to wave scattering. In this subsection, we evaluate the effectiveness of the algorithm used in this article to estimate in situ damping.

We note here that although the estimated damping ratios in this article represent small-strain damping, as all ground motions used in the inversions have PGAs less than 0.05 g, we will use the symbol $\xi$ to represent in situ damping that includes both material damping and wave scattering. $D_{min}$, on the other hand, will be used to represent small-strain damping ratio estimated from lab-based relations, such as that developed by \citet{darendeli2001development}, which exclusively captures material damping. 

\begin{figure}[htbp!]
\centering
  \subfigure[Case 1]{\includegraphics[width = 0.49\textwidth]{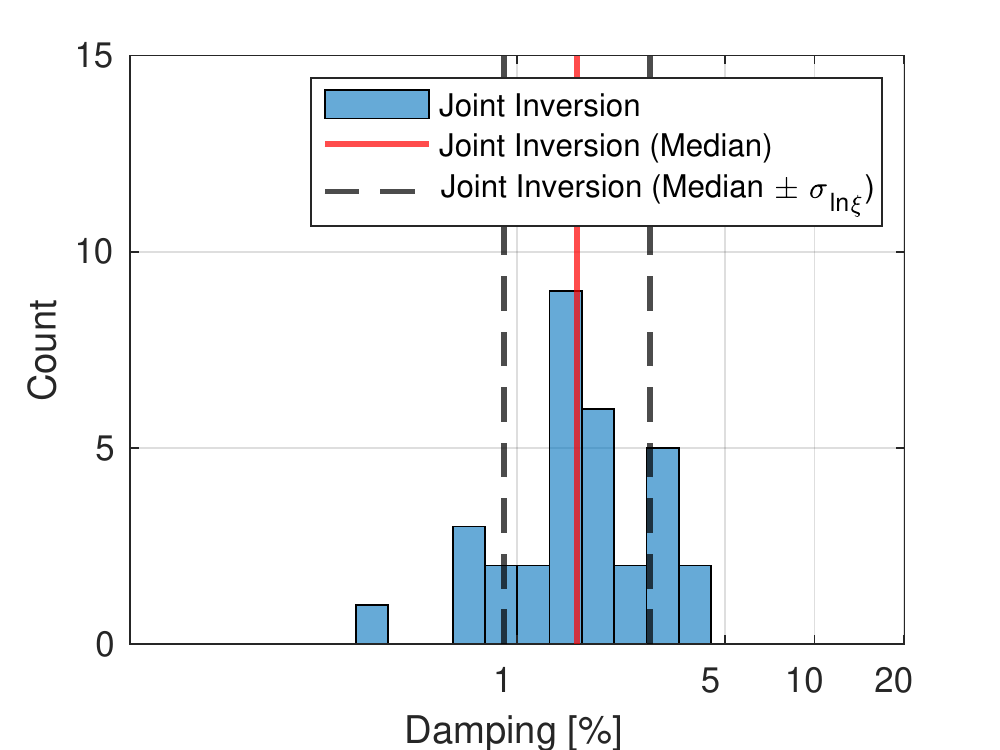}}
  \subfigure[Case 2]{\includegraphics[width = 0.49\textwidth]{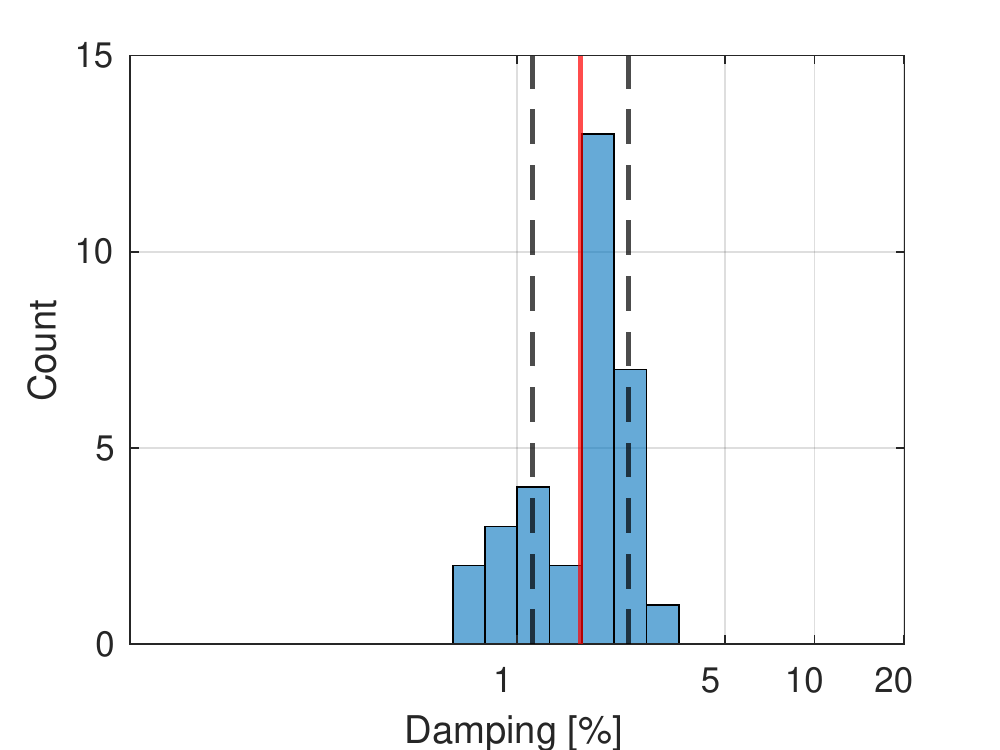}}
   \caption{Damping ratio estimation at TIDA site for Case 1 and Case 2.}
   \label{fig:TIDA-real-damping}
\end{figure}

\noindent
\underline{TIDA Site:} \Cref{fig:TIDA-real-damping} shows the distribution of estimated $\xi$ from the 32 joint inversion analyses for Cases 1 and 2 at TIDA. To compute statistics on the results, we assume a lognormal distribution. While the estimated median $\xi$ is comparable for both cases, with a value equal to approximately 1.6\%, the lognormal standard deviation of all estimated $\xi$ values ($\sigma_{ln\xi}$) is distinctly less for Case 2 (\Cref{fig:TIDA-real-damping}b), which includes prior information on the water table depth. Therefore, while it appears that including prior information on the water table might not necessarily affect the accuracy of the estimated $\xi$, it might be possible that this prior information will improve the precision of the estimation when using the algorithm adopted in this article, as evident through the lower $\sigma_{ln\xi}$.

\begin{figure}[htbp!]
\centering
  \subfigure[Case 1]{\includegraphics[width = 0.49\textwidth]{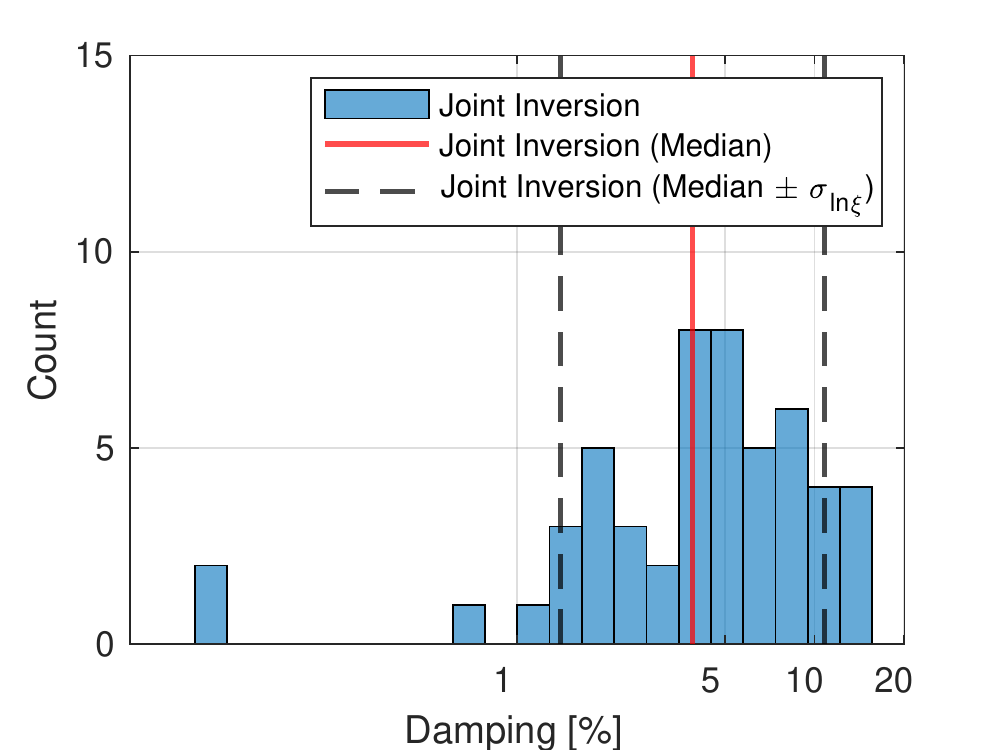}}
  \subfigure[Case 2]{\includegraphics[width = 0.49\textwidth]{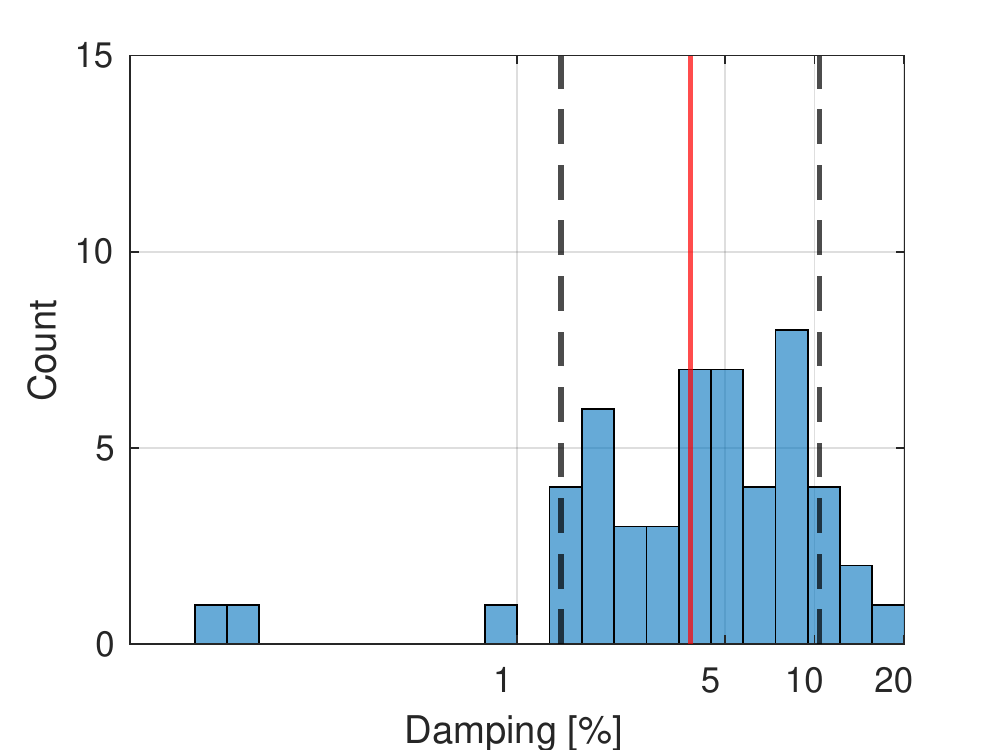}}
   \caption{Damping ratio estimation at DPDA site for Case 1 and Case 2.}
   \label{fig:DPDA-real-damping}
\end{figure}

\noindent
\underline{DPDA Site:} \Cref{fig:DPDA-real-damping} shows the distribution of estimated $\xi$ from the 52 joint inversion analyses for Cases 1 and 2 at DPDA. The median $\xi$ value from both cases are comparable and equal to approximately 3.8\%. However, unlike the results for $\sigma_{ln\xi}$ at TIDA, both cases have a comparable $\sigma_{ln\xi}$ at DPDA. Thus, including prior information on water table depth for DPDA did not affect the accuracy or precision of the estimated $\xi$. Further research is required on this topic to assess the significance, if any, of including prior information on water table depth when estimating $\xi$.

\noindent
\underline{Comparison of Results:} \Cref{table:damping-statistics} shows a summary of the $\xi$ statistics, which are also compared to the lab-based $D_{min}$ from \cite{darendeli2001development} and a modified damping meant to account for wave scattering as reported by \cite{tao2019insights}. The $D_{min}$ values from \cite{darendeli2001development} are based on estimates of in situ confining stresses obtained from typical soil unit weights and assumed plasticity indices between 10--30 for clay layers, and 0 for sand layers, similar to \cite{Hallal2021HVPart2}. Because the lab-based $D_{min}$ values vary with material properties and effective stress, and in order to compare the layer-dependent $D_{min}$ to the assumed constant damping in our estimations, we calculated a thickness-averaged $D_{min}$ for each site by simply multiplying the thickness of each layer with its $D_{min}$ and then dividing by the thickness of the full profile. This gives a single representative $D_{min}$ value that can be compared to our estimated $\xi$ values.
 
 We also compare the $\xi$ estimated in this article with those from \cite{tao2019insights}, who systematically assessed how large of an increase in $D_{min}$ is needed to account for additional damping resulting from wave scattering in the field. The authors simply multiplied the $D_{min}$ values by fixed factors between 2--6 to find the multiplier that provides the best overall fit to the ETF from recorded ground motions. They reported that a $D_{min}$ multiplicative factor of 3 is needed at TIDA, whereas a factor of 6 is needed at DPDA, to best match the median ETFs. Thus, the results in  \Cref{table:damping-statistics} for \cite{tao2019insights} are simply obtained by multiplying the $D_{min}$ values by 3 and 6 at TIDA and DPDA, respectively.
\begin{table}[htbp!]
\centering
\caption{Summary statistics of the estimated damping ratio from joint inversion for both Cases 1 and 2 at TIDA and DPDA and their comparison to average small-strain damping ratio ($D_{min}$) from \cite{darendeli2001development} and the damping reported by \cite{tao2019insights} from multiplying the $D_{min}$ by a factor between 3 and 6 to account for additional damping from wave scattering.
}
\label{table:damping-statistics}
\begin{tabular}{lclclcc}
\hline
\multirow{2}{*}{Site} & \multicolumn{2}{c}{Case 1} & \multicolumn{2}{c}{Case 2} &  Darendeli (2001) &  Tao and Rathje \\ \cline{2-5} 
                      & Median $\xi$ (\%)     &   $\sigma_{ln\xi}$  & Median $\xi$ (\%)     &   $\sigma_{ln\xi}$ & $D_{min}$ (\%) & (2019) $\xi$ (\%)    \\ \hline
TIDA                  & 1.59   & 0.57    & 1.63   & 0.37   &  0.62 & 1.86   \\
DPDA                  & 3.89   & 1.02    & 3.82   & 1.00   & 1.02  & 6.13   \\ \hline
\end{tabular}
\end{table}

The estimated median $\xi$ from joint inversion at TIDA (approximately 1.6\%) is comparable to the optimal value reported by \cite{tao2019insights}, which is equal to 1.86\%. Both values are higher than the thickness-weighted, average lab-based $D_{min}$ of 0.62\%, which provides further evidence that wave scattering contributes to the observed total damping and this effect is not accounted for in lab-based estimates. This is consistent with the findings of \cite{assimaki2007inverse} reported earlier.

While the estimated $\xi$ in this article and that from \cite{tao2019insights} are also greater than the thickness-weighted, average lab-based $D_{min}$ at DPDA, that from \cite{tao2019insights} is substantially higher. Recall that \cite{tao2019insights} reported that a multiplicative factor of 6 applied to $D_{min}$ is needed to best match the median ETF at DPDA. However, \cite{hallal2021comparison} concluded that the modified damping from \cite{tao2019insights} significantly underpredicts higher frequencies in the ETF. Therefore, the lower $\xi$ obtained in this article might be more appropriate at DPDA. This is discussed in greater detail below in the section titled \S Comparison with Other Approaches Used to Incorporate Wave Scattering into 1D GRAs.

Another interesting observation from \Cref{table:damping-statistics} is the higher estimated $\xi$ in this article at DPDA than TIDA. \cite{tao2019insights} argued that the subsurface conditions at TIDA are relatively uniform due to its hydraulic fill nature and its location in a large estuary environment, whereas those at DPDA are strongly variable due to the influence of glacial processes. Consequently, they suggested that the higher subsurface variability at DPDA leads to more wave scattering and consequently, larger apparent damping. \cite{cheng2021estimating} validated these hypotheses using a significant number of spatially distributed HVSR noise measurements at TIDA and DPDA. Their results showed more variability, over a given surface area, in the fundamental site frequency estimated from HVSR at DPDA than TIDA. Our results corroborate these findings and provide further evidence indicating that DPDA has more variability in the subsurface properties, as reflected through the higher estimated $\xi$.

\subsection{Comparison with Other 1D Approaches Used to Incorporate Wave Scattering into 1D GRAs}
\begin{figure}[htbp!]
\centering
  \subfigure[]{\includegraphics[width = 0.52\textwidth]{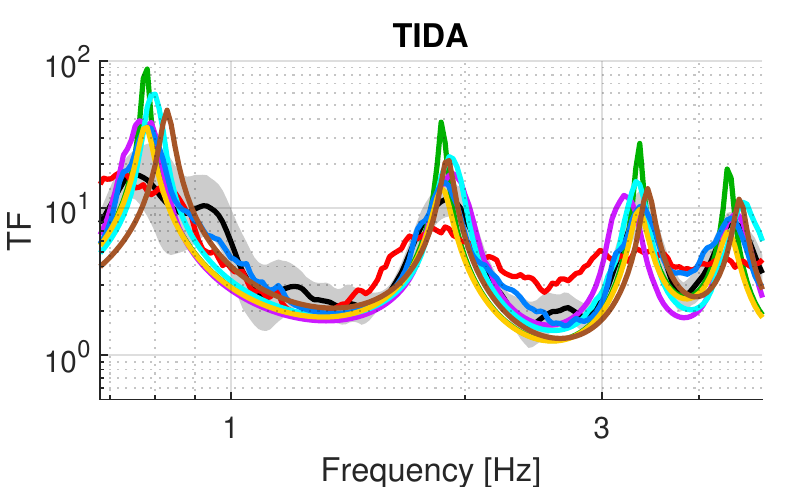}}
  \subfigure[]{\includegraphics[width = 0.455\textwidth]{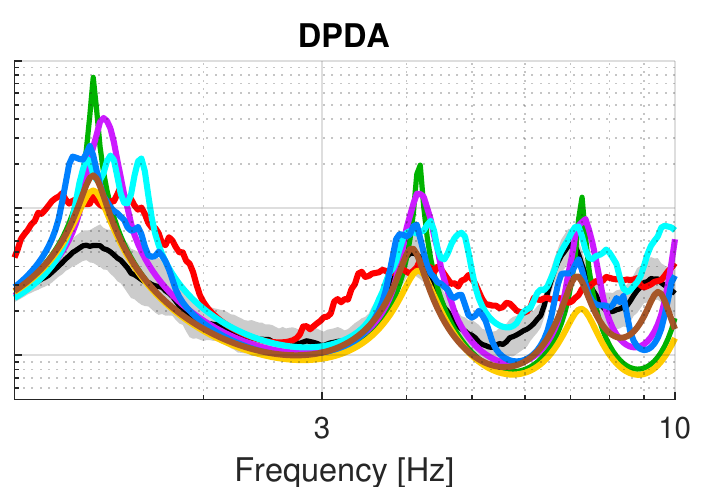}}
  {\includegraphics[width = 1.00\textwidth]{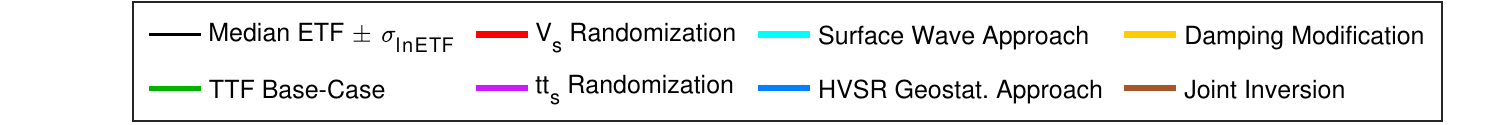}}
  \subfigure[]{\includegraphics[width = 0.52\textwidth]{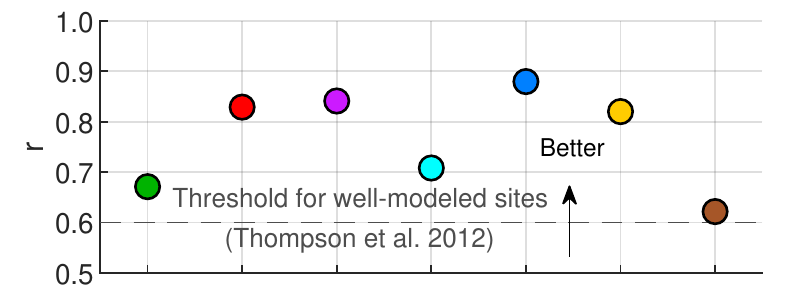}}
  \subfigure[]{\includegraphics[width = 0.455\textwidth]{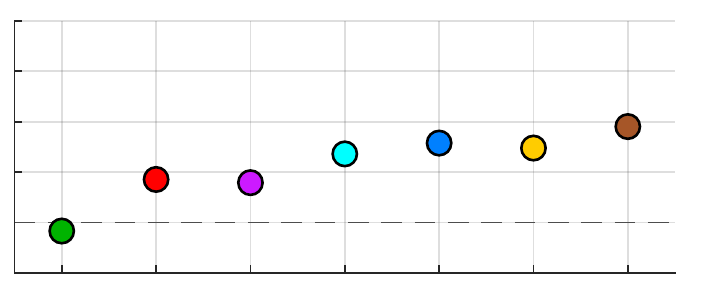}}
  \subfigure[]{\includegraphics[width = 0.52\textwidth]{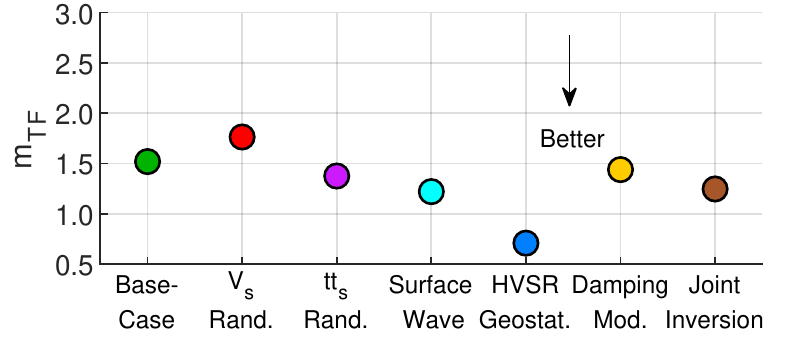}}
  \subfigure[]{\includegraphics[width = 0.455\textwidth]{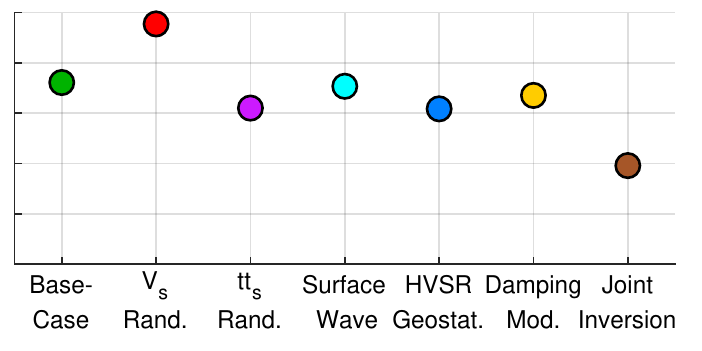}}
   \caption{Comparison of the modeling accuracy of the joint inversion results with those obtained by \cite{hallal2021comparison} at TIDA (left) and DPDA (right) from implementing alternative approaches to account for spatial variability. Shown are: (a \& b) the ETF and TTFs, (c \& d) the Pearson correlation coefficient ($r$) between the median ETF and TTFs, and (e \& f) the transfer function misfit values ($m_{TF}$) between the median ETF and TTFs at TIDA and DPDA, respectively. The transfer functions (a--b) are shown over the frequency range for which $r$ and $m_{TF}$ were computed.
   }
   \label{fig:Comparison}
\end{figure}
 In this subsection, we compare results from 1D GRAs using the material properties estimated in this article with those from other approaches used to incorporate spatial variability into 1D GRAs. \cite{hallal2021comparison} investigated five alternative approaches that can be used to account for spatial variability in 1D GRAs at TIDA and DPDA: (1) $V_s$ randomization \citep{toro1995}, (2) shear wave travel time ($tt_s$) randomization \citep{passeri2020new}, (3) utilization of suites of $V_s$ profiles derived from surface wave testing covering a large area \citep{teague2018measured}, (4) incorporation of a pseudo-3D $V_s$ model derived from an HVSR geostatistical approach \citep{Hallal2021HVPart1, Hallal2021HVPart2}, and (5) damping modifications \citep{tao2019insights}. 
 
 Similar to \cite{hallal2021comparison}, the modeling accuracy herein will be quantitatively assessed using two goodness-of-fit parameters: the Pearson correlation coefficient ($r$), which quantifies how well the shapes of the median ETF and TTF align, and the transfer function misfit ($m_{TF}$), which quantifies, on average, how many standard deviations the TTF is away from the median ETF. Note that in \S Assessment Against Experimental Data, we discussed the dispersion misfit, which is similar to $m_{TF}$, but with respect to dispersion data. \cite{thompson2012taxonomy} suggested that $r >$ 0.6 is an indicator of good alignment between TTFs and ETFs, but no recommended values exist for $m_{TF}$, partly because it depends on $\sigma_{lnETF}$, which varies from site to site. Nonetheless, lower $m_{TF}$ values are better. Both $r$ and $m_{TF}$ are computed over a frequency range in the ETF between approximately the first and fourth peaks in the ETF. Further details on the calculations can be found in \cite{hallal2021comparison}.
 
 For brevity, we will only present the final results obtained by \cite{hallal2021comparison} from implementing the five approaches at TIDA and DPDA and compare them to the 1D GRA results using the material properties estimated in this article. The results from \cite{hallal2021comparison} and their comparison to the joint inversion results from this article are presented in \Cref{fig:Comparison}. The base-case TTFs are obtained from the simplified borehole $V_s$ profiles at TIDA and DPDA (\Cref{fig:Velocity_Profiles}) with $D_{min}$ from \cite{darendeli2001development}. The TTFs for (1) $V_s$ randomization, (2) $tt_s$ randomization, (3) surface wave approach, and (4) HVSR geostatistical approach represent a mean TTF from performing numerous standard 1D GRAs using $D_{min}$ and 250 variable $V_s$ profiles for each approach. Each approach produces a different set of 250 variable $V_s$ profiles, which are meant to capture spatial variability. The TTFs for (5) damping modification are obtained from the simplified borehole $V_s$ profiles at TIDA and DPDA and the higher damping values from \cite{tao2019insights} that account for wave scattering originating from spatial discontinuities. Because the algorithm used in this article allows for calculating total damping with wave scattering effects, similar to approach (5), the TTFs for joint inversion are obtained from using only the median $V_s$ (\Cref{fig:TIDA-real-vsvp} and \Cref{fig:DPDA-real-vsvp}) and median $\xi$ values (\Cref{fig:TIDA-real-damping} and \Cref{fig:DPDA-real-damping}) estimated in this article.
 
 The joint inversion results at TIDA provide a great fit to the median ETF at higher modes, but less so at the fundamental mode. This results in a relatively unimproved $r$ value compared to the base-case TTF, and lower/poorer $r$ values than the other approaches used to incorporate spatial variability into 1D GRAs. While several researchers have highlighted the limitations of $r$ \citep{tao2019insights}, it remains widely used to judge the fit between the ETF and TTF. Conversely, the joint inversion approach yields the second best/lowest $m_{TF}$ at TIDA, only higher than the HVSR geostatistical approach, indicating a good match between the ETF and TTF. Overall, when considering both the $r$ and $m_{TF}$ values for the joint inversion approach at TIDA, which are inconsistent with one another, the results are considered as similar to all other approaches and only inferior to the HVSR geostatistical approach developed by \cite{Hallal2021HVPart1, Hallal2021HVPart2}.
 
 Interestingly, the joint inversion results at DPDA provide superior $r$ and $m_{TF}$ values compared to all approaches investigated by \cite{hallal2021comparison}. 
 The joint inversion results and those from damping modifications based on \cite{tao2019insights} provide the best reduction in the overprediction bias of the TTF at the fundamental mode. However, and as discussed earlier, the higher damping used by \cite{tao2019insights} (\Cref{table:damping-statistics}) results in more pronounced underprediciton of TTF higher mode amplification, while the joint inversion results do not significantly overdamp the higher modes. \cite{hallal2021comparison} argued all approaches that are meant to account for spatial variability are much less successful at DPDA (which is inherently more variable) compared to TIDA. However, while there is no doubt that most of the investigated approaches perform better at TIDA than DPDA, the results of this study suggest that improved 1D GRA predictions might possibly be obtained at complex sites (such as DPDA) using the joint inversion approach with proper estimation of $V_s$ and $\xi$ that accounts for the spatial variability and in situ wave scattering effects.

\section{Summary and Concluding Remarks}\label{sec:conclude}
In this article, we used a heterogeneous data assimilation technique based on the ensemble Kalman inversion to estimate $V_s$ and $V_p$ profiles and in situ damping at the TIDA and DPDA sites. The synthetic data experiments suggested that in the absence of real data noise and modeling error, joint inversion of acceleration time series and dispersion data could resolve the target $V_s$ profiles and site damping. Furthermore, we showed that the incorporation of water table depth information could help improve $V_p$ estimation, which in some cases also impacts the accuracy of $V_s$ estimation. The real data experiments also supported the importance of including this prior information. As shown, the theoretical dispersion curves and TTFs associated with estimated $V_s$ profiles and damping ratios were in good agreement with the experimental dispersion data and the ETFs. At the TIDA site, the joint inversion results' performance was inconsistently evaluated based on two goodness-of-fit parameters, but overall considered as comparable to most other methods used to incorporate 1D spatial variability into 1D GRAs. On the other hand, at the DPDA site, the joint inversion results outperformed the results of all previous studies.

As discussed in this article and in \cite{seylabi2020site}, our results suggest that the joint inversion of acceleration time series and dispersion data at downhole arrays can help improve the estimation of subsurface properties, particularly $V_s$ and $\xi$, which are critical components for all 1D GRA studies. The overall good results achieved in this article are attributed to the assimilation of acceleration time series in conjunction with dispersion data, which allow searching the parameter space to find an optimal solution. Furthermore, integrating physical constraints and a priori knowledge helped improve the well-posedness of the inverse problem at hand.

Estimating more accurate subsurface material properties at downhole array sites can be particularly important for validation studies at the Kiban-Kyoshin network (KiK-net) in Japan. Previous researchers have argued that the generally poor site response predictions at KiK-net sites are attributed to the uncertainty and low resolution of the available $V_s$ profiles \citep{afshari2019insights} and the spatial variability that cannot be captured by a single, 1D $V_s$ profiles measured in a borehole \citep{thompson2012taxonomy}. By obtaining more accurate subsurface properties using the algorithm presented in this article, researchers can draw better conclusions when preforming validation studies at downhole arrays and avoid confounding errors due to poorly estimated material properties.

The method used in this article also allows systematically incorporating wave scattering effects into damping estimation. These wave scattering effects have been shown to contribute to the total damping encountered in the field and to affect the accuracy of site response predictions. While one limitation for estimating damping using the method adopted in this article is the need for acceleration time series recorded at different depths in the subsurface, which are not available for engineering projects in practice, more similar applications at other downhole arrays will allow engineers to obtain a better understanding of the effects of wave scattering on damping. This will allow answering outstanding questions raised in recent years, such as the relation between subsurface variability and wave scattering, the extent by which lab-based damping needs to be increased to account for wave scattering in the field, and whether this increase in damping can be linked to a site-specific parameter or subsurface property. Another future development of this approach is to incorporate HVSR data in the inversion and to allow damping to vary with depth, as opposed to the depth-independent damping assumed in this article.
\bibliographystyle{apalike}
\bibliography{reference}

\begin{thebibliography}{}

\bibitem[Afshari and Stewart, 2019]{afshari2019insights}
Afshari, K. and Stewart, J.~P. (2019).
\newblock Insights from california vertical arrays on the effectiveness of
  ground response analysis with alternative damping models.
\newblock {\em Bulletin of the Seismological Society of America},
  109(4):1250--1264.

\bibitem[Albers et~al., 2019]{albers2019ensemble}
Albers, D.~J., Blancquart, P.-A., Levine, M.~E., Seylabi, E.~E., and Stuart,
  A.~M. (2019).
\newblock Ensemble kalman methods with constraints.
\newblock {\em Inverse Problems,
  \url{http://iopscience.iop.org/10.1088/1361-6420/ab1c09}}.

\bibitem[Assimaki and Steidl, 2007]{assimaki2007inverse}
Assimaki, D. and Steidl, J. (2007).
\newblock Inverse analysis of weak and strong motion downhole array data from
  the mw7. 0 sanriku-minami earthquake.
\newblock {\em Soil Dynamics and Earthquake Engineering}, 27(1):73--92.

\bibitem[Chen and Oliver, 2012]{O1}
Chen, Y. and Oliver, D.~S. (2012).
\newblock Ensemble randomized maximum likelihood method as an iterative
  ensemble smoother.
\newblock {\em Mathematical Geosciences}, 44(1):1--26.

\bibitem[Cheng et~al., 2021]{cheng2021estimating}
Cheng, T., Hallal, M.~M., Vantassel, J.~P., and Cox, B.~R. (2021).
\newblock Estimating unbiased statistics for fundamental site frequency using
  spatially distributed hvsr measurements and voronoi tessellation.
\newblock {\em Journal of Geotechnical and Geoenvironmental Engineering},
  147(8):04021068.

\bibitem[Combellick, 1999]{combellick1999simplified}
Combellick, R.~A. (1999).
\newblock Simplified geologic map and cross sections of central and east
  anchorage, alaska.
\newblock Technical report, Division of Geological \& Geophysical Surveys
  Preliminary Interpretive Rep. No. 1999-1, Fairbanks, Alaska.

\bibitem[Cox and Teague, 2016]{cox2016layering}
Cox, B.~R. and Teague, D.~P. (2016).
\newblock Layering ratios: a systematic approach to the inversion of surface
  wave data in the absence of a priori information.
\newblock {\em Geophysical Journal International}, 207(1):422--438.

\bibitem[Darendeli, 2001]{darendeli2001development}
Darendeli, M.~B. (2001).
\newblock {\em Development of a new family of normalized modulus reduction and
  material damping curves}.
\newblock PhD thesis, University of Texas at Austin.

\bibitem[Dunkin, 1965]{dunkin1965computation}
Dunkin, J.~W. (1965).
\newblock Computation of modal solutions in layered, elastic media at high
  frequencies.
\newblock {\em Bulletin of the Seismological Society of America},
  55(2):335--358.

\bibitem[Emerick and Reynolds, 2013]{O2}
Emerick, A.~A. and Reynolds, A.~C. (2013).
\newblock Investigation of the sampling performance of ensemble-based methods
  with a simple reservoir model.
\newblock {\em Computational Geosciences}, 17(2):325--350.

\bibitem[Evensen, 2009]{evensen2009data}
Evensen, G. (2009).
\newblock {\em Data assimilation: the ensemble Kalman filter}.
\newblock Springer Science \& Business Media.

\bibitem[Foerster and Modaressi, 2007]{foerster2007nonlinear}
Foerster, E. and Modaressi, H. (2007).
\newblock Nonlinear numerical method for earthquake site response analysis ii
  — case studies.
\newblock {\em Bulletin of Earthquake Engineering}, 5(3):325--345.

\bibitem[Gibbs et~al., 1992]{gibbs1992seismic}
Gibbs, J.~F., Fumal, T.~E., Boore, D.~M., and Joyner, W.~B. (1992).
\newblock Seismic velocities and geologic logs from borehole measurements at
  seven strong-motion stations that recorded the 1989 loma prieta earthquake.
\newblock Technical report, US Geological Survey Open-File Report 92-287, Menlo
  Park, California.

\bibitem[Goulet et~al., 2014]{goulet2014peer}
Goulet, C.~A., Kishida, T., Ancheta, T.~D., Cramer, C.~H., Darragh, R.~B.,
  Silva, W.~J., Hashash, Y.~M., Harmon, J., Stewart, J.~P., Wooddell, K.~E.,
  and Youngs, R.~R. (2014).
\newblock Peer nga-east database.
\newblock {\em PEER Report 2014}, 17.

\bibitem[Graizer and Shakal, 2004]{graizer2004analysis}
Graizer, V. and Shakal, A. (2004).
\newblock Analysis of some of csmip strong-motion geotechnical array
  recordings.
\newblock In {\em Proceedings of the International Workshop for Site Selection,
  Installation and Operation of Geotechnical Strong-Motion Arrays: Workshop},
  volume~1.

\bibitem[Hallal and Cox, 2021a]{Hallal2021HVPart1}
Hallal, M.~M. and Cox, B.~R. (2021a).
\newblock An h/v geostatistical approach for building pseudo-3d vs models to
  account for spatial variability in ground response analyses part i: Model
  development.
\newblock {\em Earthquake Spectra}, page 8755293020981989.

\bibitem[Hallal and Cox, 2021b]{Hallal2021HVPart2}
Hallal, M.~M. and Cox, B.~R. (2021b).
\newblock An h/v geostatistical approach for building pseudo-3d vs models to
  account for spatial variability in ground response analyses part ii:
  Application to 1d analyses at two downhole array sites.
\newblock {\em Earthquake Spectra}.

\bibitem[Hallal et~al., 2021]{hallal2021comparison}
Hallal, M.~M., Cox, B.~R., and Vantassel, J.~P. (2021).
\newblock Comparison of state-of-the-art approaches used to account for spatial
  variability in 1d ground response analyses.
\newblock {\em Journal of Geotechnical and Geoenvironmental Engineering}.

\bibitem[Haskell, 1953]{haskell1953dispersion}
Haskell, N.~A. (1953).
\newblock The dispersion of surface waves on multilayered media.
\newblock {\em Bulletin of the seismological Society of America}, 43(1):17--34.

\bibitem[Iglesias et~al., 2013]{iglesias2013ensemble}
Iglesias, M.~A., Law, K.~J., and Stuart, A.~M. (2013).
\newblock Ensemble kalman methods for inverse problems.
\newblock {\em Inverse Problems}, 29(4):045001.

\bibitem[Knopoff, 1964]{knopoff1964matrix}
Knopoff, L. (1964).
\newblock A matrix method for elastic wave problems.
\newblock {\em Bulletin of the Seismological Society of America},
  54(1):431--438.

\bibitem[Kramer, 1996]{Kramer_1996_Book}
Kramer, S.~L. (1996).
\newblock {\em Geotechnical Earthquake Engineering}.
\newblock International Series in Civil Engineering and Engineering Mechanics.
  Prentice Hall, Upper Saddle River, New Jersey, 1st edition.

\bibitem[Pass, 1994]{pass1994soil}
Pass, D.~G. (1994).
\newblock Soil characterization of the deep accelerometer site at treasure
  island, san francisco, california.
\newblock Master's thesis, University of New Hampshire, USA.

\bibitem[Passeri et~al., 2020]{passeri2020new}
Passeri, F., Foti, S., and Rodriguez-Marek, A. (2020).
\newblock A new geostatistical model for shear wave velocity profiles.
\newblock {\em Soil Dynamics and Earthquake Engineering}, 136:106247.

\bibitem[Phillips and Hashash, 2009]{phillips2009damping}
Phillips, C. and Hashash, Y.~M. (2009).
\newblock Damping formulation for nonlinear 1d site response analyses.
\newblock {\em Soil Dynamics and Earthquake Engineering}, 29(7):1143--1158.

\bibitem[Pilz and Cotton, 2019]{pilz2019does}
Pilz, M. and Cotton, F. (2019).
\newblock Does the one-dimensional assumption hold for site response analysis?
  a study of seismic site responses and implication for ground motion
  assessment using kik-net strong-motion data.
\newblock {\em Earthquake Spectra}, 35(2):883--905.

\bibitem[Rollins et~al., 1994]{rollins1994ground}
Rollins, K.~M., Mchood, M.~D., Hryciw, R.~D., Homolka, M., and Shewbridge,
  S.~E. (1994).
\newblock Ground response on treasure island.
\newblock Technical report, U.S. Geological Survey Professional Paper 1551 A.

\bibitem[Seylabi et~al., 2020]{seylabi2020site}
Seylabi, E., Stuart, A.~M., and Asimaki, D. (2020).
\newblock Site characterization at downhole arrays by joint inversion of
  dispersion data and acceleration time series.
\newblock {\em Bulletin of the Seismological Society of America},
  110(3):1323--1337.

\bibitem[Tao and Rathje, 2019]{tao2019insights}
Tao, Y. and Rathje, E. (2019).
\newblock Insights into modeling small-strain site response derived from
  downhole array data.
\newblock {\em Journal of Geotechnical and Geoenvironmental Engineering},
  145(7):04019023.

\bibitem[Teague et~al., 2018]{teague2018measured}
Teague, D.~P., Cox, B.~R., and Rathje, E.~M. (2018).
\newblock Measured vs. predicted site response at the garner valley downhole
  array considering shear wave velocity uncertainty from borehole and surface
  wave methods.
\newblock {\em Soil Dynamics and Earthquake Engineering}, 113:339--355.

\bibitem[Thompson et~al., 2012]{thompson2012taxonomy}
Thompson, E.~M., Baise, L.~G., Tanaka, Y., and Kayen, R.~E. (2012).
\newblock A taxonomy of site response complexity.
\newblock {\em Soil Dynamics and Earthquake Engineering}, 41:32--43.

\bibitem[Thomson, 1950]{thomson1950transmission}
Thomson, W.~T. (1950).
\newblock Transmission of elastic waves through a stratified solid medium.
\newblock {\em Journal of applied Physics}, 21(2):89--93.

\bibitem[Thornley et~al., 2019]{thornley2019situ}
Thornley, J., Dutta, U., Fahringer, P., and Yang, Z. (2019).
\newblock In situ shear-wave velocity measurements at the delaney park downhole
  array, anchorage, alaska.
\newblock {\em Seismological Research Letters}, 90(1):395--400.

\bibitem[Toro, 1995]{toro1995}
Toro, G. (1995).
\newblock {\em Probabilistic models of site velocity profiles for generic and
  site-specific ground-motion amplification studies}.
\newblock Upton, NY: Brookhaven National Laboratory.

\bibitem[Wathelet, 2005]{wathelet2005array}
Wathelet, M. (2005).
\newblock Array recordings of ambient vibrations: surface-wave inversion.
\newblock {\em PhD Diss., Li{\'e}ge University}, 161.

\bibitem[Wathelet et~al., 2004]{wathelet2004surface}
Wathelet, M., Jongmans, D., and Ohrnberger, M. (2004).
\newblock Surface-wave inversion using a direct search algorithm and its
  application to ambient vibration measurements.
\newblock {\em Near surface geophysics}, 2(4):211--221.

\end{thebibliography}
\end{document}